\documentclass{CSML}

\pdfoutput=1

\usepackage{lastpage}

\def\dOi{13(4:22)2017}
\lmcsheading%
{\dOi}
{1--\pageref{LastPage}}
{}
{}
{Jun.~\phantom01, 2016}
{Nov.~29, 2017}
{}

\usepackage[utf8]{inputenc}

\usepackage{color}
\usepackage[hidelinks]{hyperref}
\usepackage{float}
\usepackage{amsmath}
\usepackage{amsfonts}
\usepackage{amsthm}
\usepackage{amssymb}
\usepackage{proof}
\usepackage{mathpartir}
\usepackage{mathrsfs}
\usepackage{stmaryrd}
\usepackage{cmll}
\usepackage{graphicx}
\usepackage[all]{xy}
\usepackage{listings}
\usepackage{todonotes}
\usepackage{Guyboxes}

\DeclareMathOperator{\domain}{dom}
\DeclareMathOperator{\codomain}{cod}
\DeclareMathOperator{\cod}{cod}
\DeclareMathOperator{\dom}{dom}
\DeclareMathOperator{\ctxof}{ctx-of}
\DeclareMathOperator{\typeof}{type-of}

\newcommand{\vdashS}{\vdash}
\newcommand {\emptyContext}{1}

\newcommand {\q}[2]{{\tt q}_{#1, #2}}
\newcommand {\qI}{{\tt q}}

\newcommand{\emptySub}[1]{\emptySubI_{#1}}
\newcommand{\emptySubI}{\langle\rangle}

\newcommand{\Ctx}{\mathrm{Ctx}}
\newcommand{\Sub}{\mathrm{Sub}}
\newcommand{\Ty}{\mathrm{Ty}}
\newcommand{\Tm}{\mathrm{Tm}}
\newcommand{\C}{{\mathcal C}}
\newcommand{\I}{{\mathcal I}}
\newcommand{\T}{{\mathcal T}}

\newcommand{\arrow}{{\rightarrow}}
\newcommand{\RawCtx}{{\tt Ctx}}
\newcommand{\RawSub}{{\tt Sub}}
\newcommand{\RawTy}{{\tt Ty}}
\newcommand{\RawTm}{{\tt Tm}}

\newcommand{\inte}[1]{\llbracket #1 \rrbracket}
\newcommand{\intCtx}[1]{\llbracket #1 \rrbracket}

\newcommand{\iniCtx}[1]{\overline{\llbracket #1 \rrbracket}}
\newcommand{\iniSub}[3]{\overline{\llbracket #3 \rrbracket}_{#1,#2}}
\newcommand{\iniTy}[2]{\overline{\llbracket #2 \rrbracket}_{#1}}
\newcommand{\iniTm}[3]{\overline{\llbracket #3 \rrbracket}_{#1,#2}}

\newcommand{\refl}[0]{{\rm ref}}

\newcommand{\ext}[1]{\langle #1 \rangle}

\newcommand{\omitthis}[1]{}

\newcommand{\changenote}[1]{}

 \newcommand{\Id}[0]{{\rm I}}

\newcommand{\longtext}[1]{}
\newcommand{\shorttext}[1]{}
\newcommand{\commentaway}[1]{}

\definecolor{Red}{rgb}{1,0,0}

\definecolor{White}{rgb}{1,1,1}
\newcommand{\white}[1]{{\color{White}#1}}

\renewcommand{\bar}[1]{\overline{#1}}

\newdir{pb}{:(1,-1)@^{|-}}
\def\pb#1{\save[]+<16 pt,0 pt>:a(#1)\ar@{pb{}}[]\restore}

\newcommand{\Fam}{\textbf{Fam}}
\newcommand{\nilc}{1}
\newcommand{\cext}{.}
\newcommand{\indexed}[1]{\boldsymbol{#1}}
\newcommand{\Cat}{\mathrm{Cat}}
\newcommand{\op}{\text{op}}
\newcommand{\iso}{\cong}
\newcommand{\subst}[1]{\langle #1 \rangle}
\newcommand{\applyopen}[2]{\{ #1 \}  #2 }

\def\N{\mathrm{N}}
\def\U{\mathrm{U}}
\def\p{{\tt p}}
\def\ev{{\tt ev}}
\def\q0{{\tt q}}

\def\arrow{\rightarrow}

\def\GammaCL{\Gamma_{\mathrm{CL}}}

\def\I{\mathrm{I}}
\def\refl{\mathrm{r}}
\def\id{{\tt id}}

\def\idC{\mathrm{id}_\C}
\newcommand{\pair}{\mathrm{pair}}
\newcommand{\fst}{\mathrm{fst}}
\newcommand{\interp}[1]{ \overline{\llbracket #1 \rrbracket}}
\newcommand{\Cwf}{\textbf{CwF}}
\newcommand{\Cwfs}{\Cwf_s}
\newcommand{\D}{\mathcal{D}}
\newcommand{\snd}{\mathrm{snd}}
\newcommand{\ap}{\mathrm{app}}
\newcommand{\app}{\mathrm{app}}

\newcommand{\TT}{\mathbf{T}}

\title[Undecidability of Equality in the Free Locally Cartesian Closed Category]{Undecidability of Equality\\ in the Free Locally Cartesian Closed Category (extended version)\rsuper*}
\author[S.~Castellan]{Simon Castellan\rsuper a}
\address{{\lsuper{a,b}}Univ Lyon, CNRS, ENS de Lyon, UCB Lyon 1, LIP}
\email{\{simon.castellan,pierre.clairambault\}@ens-lyon.fr}
\author[P.~~Clairambault]{Pierre Clairambault\rsuper b}
\author[P.~Dybjer]{Peter Dybjer\rsuper c}
\address{{\lsuper c}Chalmers University of Technology}
\email{peterd@chalmers.se}

\begin{document}

\titlecomment{{\lsuper*}This paper is an extended version of \cite{DBLP:conf/tlca/CastellanCD15}}

\maketitle

\begin{abstract}
  We show that a version of Martin-Löf type theory with an extensional
  identity type former $\I$, a unit type $\N_1$, $\Sigma$-types, $\Pi$-types, and a base type is a free
  category with families (supporting these type formers) both in a 1-
  and a 2-categorical sense. It follows that the underlying category
  of contexts is a free locally cartesian closed category in a
  2-categorical sense because of a previously proved biequivalence. We
  show that equality in this category is undecidable by reducing
  it to the undecidability of convertibility in combinatory logic. Essentially the same construction also shows a slightly strengthened form of the result that equality in extensional Martin-Löf type theory with one universe is undecidable.
\end{abstract}

\allowdisplaybreaks
\section{Introduction}

In previous work \cite{clairambault:novisad,clairambault:mscs} we showed the biequivalence of locally cartesian closed categories (lcccs) and the $\I, \Sigma, \Pi$-fragment of extensional Martin-Löf type theory. More precisely, we showed the biequivalence of the following two 2-categories.
\begin{itemize}
\item
The first has as {\em objects} lcccs, as {\em arrows} functors which preserve the lccc-structure (up to isomorphism), and as {\em 2-cells} natural transformations.
\item
The second has as {\em objects} categories with families (cwfs)
\cite{dybjer:torino,hofmann:cambridge}
which support extensional identity types ($\I$-types), $\Sigma$-types, $\Pi$-types, and are {\em democratic}, as {\em arrows} pseudo cwf-morphisms (preserving structure up to isomorphism), and as {\em 2-cells} pseudo cwf-transformations. A cwf is democratic iff there is an equivalence between its category of contexts and its category of closed types. 
\end{itemize}
This result is a corrected version of a result by Seely \cite{seely:lccc} concerning the equivalence of the category of lcccs and the category of Martin-Löf type theories. Seely's paper did not address the coherence problem caused  by the interpretation of substitution as pullbacks \cite{CurienPL:subi}. As Hofmann showed \cite{HofmannM:intttl}, this coherence problem can be solved by extending a construction of B\'enabou \cite{benabou}. Our biequivalence is based on this construction.

Cwfs are models of the most basic rules of dependent type theory; those dealing with substitution, assumption, and context formation, the rules which come before any rules for specific type formers.
The distinguishing feature of cwfs, compared to other categorical notions of model of dependent types, is that they are formulated in a way which makes the connection with the ordinary syntactic formulation of dependent type theory transparent. They can be defined purely equationally \cite{dybjer:torino} as a generalised algebraic theory (gat) \cite{cartmell:apal}, where each sort symbol corresponds to a judgment form, and each operator symbol corresponds to an inference rule in a variable free formulation of Martin-Löf's explicit substitution calculus for dependent type theory \cite{martinlof:gbg92,tasistro:lic}.

Cwfs provide a basic theory of dependently typed $n$-place functions. We remark that {\em non-dependent} cwfs, in which there is a {\em fixed} set of {\em types}, are closely related to (cartesian) multicategories, where the {\em terms} of the cwf correspond to {\em multiarrows}. A difference is however that a multiarrow always comes with a finite list of input objects, whereas the cwf-axioms do not force the input context of a term to be a list.

Cwfs are not only models of dependent type theory, but also suggest an answer to the question what dependent type theory is as a mathematical object. Perhaps surprisingly, this is a non-trivial question, and Voevodsky has remarked that ``a type system is not a mathematical notion''. There are numerous variations of Martin-Löf type theory in the literature, even of the formulation of the most basic rules for dependent types. There are systems with explicit and implicit substitutions, and there are variations in assumption, context formation, and substitution rules. There are formulations with de Bruijn indices and with ordinary named variables, etc. In fact, there are so many rules that most papers do not try to provide a complete list; and if you do try to list all of them how can you be sure that you have not forgotten any? Nevertheless, there is a tacit assumption that most variations are equivalent and that a complete list of rules could be given if needed. However, from a mathematical point of view this is neither clear nor elegant. 

To remedy this situation we suggest to define Martin-Löf type theory (and other dependent type theories) abstractly as the initial cwf (with extra structure). The category of cwfs and morphisms which preserve cwf-structure on the nose was defined by Dybjer \cite{dybjer:torino}. We suggest that the correctness of a definition or an implementation of dependent type theory means that it gives rise to an initial object in this category of cwfs (with extra structure). Here we shall construct the initial object in this category explicitly in the simplest possible way
following closely the definition of the generalised algebraic theory of cwfs. Note however that the notion of a generalised algebraic theory is itself based on dependent type theory, that is, on cwf-structure. So just defining the initial cwf as the generalised algebraic theory of cwfs would be circular. 

Instead we construct the initial cwf explicitly by giving grammar and inference rules which follow
closely the operators of the gat of cwfs. However, we must also make equality reasoning explicit. To
decrease the number of rules, we present a ``per-style'' system rather than an ordinary one. We will
mutually define four partial equivalence relations (pers): for the judgments of context equality
$\Gamma = \Gamma'$, substitution equality $\Delta \vdash \gamma = \gamma' : \Gamma$, type equality
$\Gamma \vdash A = A'$, and term equality  $\Gamma \vdash a = a' : A$. The ordinary judgments will
be defined as the reflexive instances. For example, $\Gamma \vdash a : A$ will be defined as $\Gamma
\vdash a = a : A$. There are altogether 32 inference rules for the pure theory of cwfs: the first 8 rules express that we define four families of pers; the second 3 rules that judgments preserve equality of contexts and types; the
next 10 rules express the typing and congruence of the 10 cwf-operations; and the final 11 rules are
the conversion rules for these operations. In addition to the pure theory of cwfs, we have 1 rule for the base type.

Our only optimisation is the elimination of some redundant arguments of operators. For example, the composition operator in the gat of cwfs has five arguments: three objects and two arrows. However, the three object arguments can be recovered from the arrows, and can hence be omitted. 

The goal of the present paper is to prove the undecidability of
equality in the free lccc. To this end we extend our formal system for
cwfs with rules for extensional $\I$-types, $\N_1, \Sigma, \Pi,$ and a
base type.  (Note that we have added the unit type $\N_1$ to the type
formers needed for the proof of biequivalence with lcccs. This is
because we need to construct a democratic cwf, where there is a
bijection between types and contexts (see above). Therefore we need
the type $\N_1$ which corresponds to the empty context.) There are 5
rules for $\I$-types, 3 rules for $\N_1$, 11 rules for $\Sigma$, and 8
rules for $\Pi$. We want to show that this yields a free lccc on one
object, by appealing to our biequivalence theorem. However, in order
to use our biequivalence it does not suffice to show that we get a
free cwf in the 1-category of cwfs and strict cwf-morphisms: we must
show that it is also free (``bifree'') in the 2-category of cwfs and
pseudo cwf-morphisms. Indeed, biequivalences do not preserve
  initial objects in general: uniquess of a morphism $\textbf{0} \rightarrow A$
  out of an initial object is lost.  The proof of bifreeness is technically more involved
because of the complexity of the notion of pseudo cwf-morphism.

Once we have constructed the free lccc (as a cwf-formulation of Martin-Löf type theory with extensional $\I$-types, $\N_1, \Sigma, \Pi$, and one base type) we will be able to prove undecidability. 
It is well-known that extensional Martin-L\"of type theory with one universe (folklore) or with
natural numbers \cite{hofmann:thesis} has undecidable equality, and we only need to show that a
similar construction can be made without a universe and without natural numbers, provided we have a
base type. We do this by encoding untyped combinatory logic as a context, and use the undecidability
of equality in this theory.

\medskip

\noindent {\bf Related work.} Palmgren and Vickers \cite{palmgren:apal} show how to construct free models of essentially algebraic theories in general. We could use this result to build a free cwf, but this only shows freeness in the 1-categorical sense. We also think that the explicit construction of the free (and bifree) cwf is interesting in its own right.

\medskip

\noindent {\bf Plan.} In Section 2 we prove a few undecidability theorems, including the undecidability of equality in Martin-Löf type theory with extensional $\I$-types, $\Pi$-types, and one base type. In Section 3 we construct a free cwf on one base type. We show that it is free and bifree. In Section 4 we construct a free and bifree cwf with extensional identity types, $\N_1, \Sigma, \Pi$, and one base type. Since this cwf is democratic we can use the biequivalence result to conclude that this yields a free lccc in a 2-categorical sense.

\section{Undecidability in Martin-Löf type theory}

Like any other single-sorted first order equational theory, combinatory logic can be encoded as a context
in Martin-Löf type
theory with $\I$-types, $\Pi$-types, and a base type $o$. The
context $\GammaCL$ for combinatory logic is the following:\newpage
{\white x}\vspace{-24 pt}\begin{eqnarray*}
k &:& o,\\
s &:& o,\\
\cdot  &:& o \arrow o \arrow o,\\
ax_{k} &:& \Pi x y : o.\ \I(o,\,k\cdot x\cdot y,\, x),\\
ax_{s} &:& \Pi x y z : o.\ \I(o,\, s\cdot x\cdot y\cdot z,\, x\cdot z\cdot (y\cdot z))
\end{eqnarray*}
The left-associative binary infix symbol ``$\cdot$'' stands for
application. Note that $k,s,\cdot ,ax_k, ax_s$ are all variables.

\begin{thm}
Type-inhabitation in  Martin-Löf type
theory with (intensional or extensional) identity-types, $\Pi$-types and a base type is undecidable. 
\end{thm}
This follows from the undecidability of convertibility in combinatory logic, since the type
$$
\GammaCL \vdash \I(o,\, M,\, M')
$$
is inhabited iff the closed combinatory terms $M$ and $M'$ are
convertible. Clearly, if the combinatory terms are convertible, it 
can be formalised in this fragment of type theory. For the other direction we build a
model of the context $\GammaCL$ where $o$ is interpreted as the set of
combinatory terms modulo convertibility.

\begin{thm}\label{thm:undecidable_tt}
Judgmental equality in Martin-Löf type
theory with extensional identity-types, $\Pi$-types and a base type is undecidable. 
\end{thm}
With extensional identity types \cite{martinlof:hannover} the above identity type is inhabited iff
the corresponding equality judgment is valid:
$$
\GammaCL \vdash M = M' : o
$$

This theorem also holds if we add $\N_1$ and $\Sigma$-types to the theory. The remainder of the paper will show that the category of contexts of the resulting fragment of Martin-Löf type theory is bifree in the 2-category of lcccs (Theorem \ref{bifree-lccc}). Our main result follows:
\begin{thm}
Equality of arrows in the bifree lccc on one object is undecidable.
\end{thm}
We remark that the following folklore theorem can be proved in the same way as Theorem
\ref{thm:undecidable_tt}. (We are not aware of
any published proof of this theorem, but see Hofmann \cite{hofmann:thesis} for a proof which instead
uses the natural number type.)
\begin{thm}
Judgmental equality in Martin-Löf type
theory with extensional identity-types, $\Pi$-types and a universe $\U$ is undecidable. 
\end{thm}
If we have a universe we can instead work in the context
\begin{eqnarray*}
X &:& \U,\\
k &:& X,\\
s &:& X,\\
\cdot  &:& X \arrow X \arrow X,\\
ax_{k} &:& \Pi x y : X.\ \I(X,\, k\cdot x\cdot y,\, x),\\
ax_{s} &:& \Pi x y z : X.\ \I(X,\, s\cdot x\cdot y\cdot z,\, x\cdot z\cdot (y\cdot z))
\end{eqnarray*}
and prove undecidability for this theory (without a base type) in the same way as above.

Note that we don't need any closure properties at all for $\U$ -- only the ability to quantify over small types. Hence we prove a slightly stronger theorem than the folklore theorem which assumes that $\U$ is closed under function types and uses the context
\begin{eqnarray*}
X &:& \U,\\
x &:& \I(\U,X, X \arrow X)
\end{eqnarray*}
so that $X$ is a model of the untyped lambda calculus. 

\section{A free category with families}
\label{sec:freecwf}

In this section we define a free cwf syntactically, as a {\em term
  model} consisting of derivable well-formed contexts, substitutions, types and
terms modulo derivable equality. To this end we give syntax and
inference rules for a cwf-calculus, that is, a variable-free explicit
substitution calculus for dependent type theory. 

We first prove that this calculus yields a free cwf in the category
where morphisms preserve cwf-structure on the nose. The free cwf on one object is a rather degenerate structure, since there are no non-trivial dependent types. However, we have nevertheless chosen to present this part of the construction separately. Cwfs model the common core of dependent type theory, including all generalised algebraic theories, pure type systems \cite{barendregt:tlc}, and fragments of Martin-Löf type theory. The construction of a free pure cwf is thus the common basis for constructing free and initial cwfs with appropriate extra structure for modelling specific dependent type theories.

In Section \ref{sec:cwfs} we  start by recalling the definition of cwfs, the associated
morphisms -- both those preserving structure in the strict sense and up to isomorphism -- and some
related definitions and notations. In Section \ref{sec:syntax}, we  introduce our syntax and
inference rules. In Section \ref{sec:freenessT}, we  show that these inference rules give rise to
a \emph{free} cwf, in the category of cwfs and strict cwf-morphisms. Finally, in Section \ref{a-free-cwf} we
prove that our free cwf is also bifree in the $2$-category of cwf-morphisms preserving structure up
to isomorphism.


\subsection{The 2-category of categories with families}
\label{sec:cwfs}

The 2-category of cwfs and pseudo-morphisms which preserve
cwf-structure up to isomorphism was defined in
\cite{clairambault:novisad,clairambault:mscs}. Here we only give an
outline.

\medskip

\noindent{\bf Notations.} We write $\textbf{Fam}$ for the category of
families of sets: objects are families of sets $(X_i)_{i \in I}$ and
maps from $(X_i)_{i \in I}$ to $(Y_j)_{j \in J}$ are pairs
$(f : I \rightarrow J, (f_i : X_i \rightarrow Y_{f(i)})_{i \in
  I})$.
In a category with families, contexts and substitutions form the objects and arrows of a category $\C$. The set of objects will be written $\Ctx_\C$
and the set of morphisms from $ \Delta $ to $ \Gamma $ will be written
$\Sub_\C( \Delta , \Gamma )$. Types and terms over a context
$ \Gamma $ form a family
$(\Tm_\C( \Gamma , A))_{A \in \Ty_\C\Gamma }$, and substitution gives rise to
a functorial action on such a family. Thus we have a functor
\[
T : \C^{\text{op}} \rightarrow \textbf{Fam}
\] 

The action of $T$ on objects is $T \Gamma = (\Tm_\C( \Gamma , A))_{A \in \Ty_\C \Gamma
}$, and its action on a type $A$ is written $A[\_]$: if
$ \gamma \in \Sub_\C(\Gamma,\Delta)$ and $A \in \Ty_\C(\Delta)$, then
$A[\gamma] \in \Ty_\C(\Gamma)$.  Similarly, if
$a \in \Tm_\C( \Delta , A)$, we write
$a[\gamma] \in \Tm_\C ( \Gamma , A[ \gamma ])$ for the functorial
action of $T$ on $a$.


\begin{defi}[Category with families]
  A cwf is given by a category $\C$ and a functor $T : \C^{\text{op}} \to \textbf{Fam}$
together with the following chosen structure:\newpage

  \begin{itemize}
  \item (\emph{Empty context}) $\C$ has a terminal object $1$.
  \item (\emph{Context comprehension}) For each $ \Delta \in \Ctx_\C$ and
    $A \in \Ty_\C( \Delta )$ there is the extended context $ \Delta .A \in \Ctx_\C$
    with a substitution $\p_{A} \in \Sub_\C(\Delta .A ,\Delta) $ and a term
    $\qI_A \in \Tm_\C( \Delta .A, A [\p_A])$, such that for every pair
    $ \gamma \in \Sub_\C(\Gamma,\Delta)$ and
    $a \in \Tm_\C( \Gamma , A [\gamma] )$ there exists a unique
\[
\langle \gamma, a \rangle \in \Sub_\C(\Gamma, \Delta .A)
\]
 such
    that $\p_A \circ \langle \gamma , a \rangle = \gamma $ and
    $\qI_A [\langle \gamma , a \rangle] = a $.
  \end{itemize}

\end{defi}

\noindent Note that with the notation $\Ty_\C$ and $\Tm_\C$ there is no need to
explicitly mention the functor $T$ when working with categories with
families, and we will often keep it implicit. Given a substitution
$ \gamma : \Delta \rightarrow \Gamma $, and $A \in \Ty_\C( \Gamma )$,
we write $ \gamma \uparrow A$ or $ \gamma ^+$ (when $A$ can be
inferred from the context) for the lifting of $ \gamma $ to $A$:
$ \langle \gamma \circ \p, \qI \rangle : \Delta .A[\gamma] \rightarrow
\Gamma .A$.

\medskip

\noindent\textbf{The indexed category.}  In
\cite{clairambault:novisad,clairambault:mscs} it is shown that any
cwf $\C$ induces a functor
$\TT: \C^{\text{op}} \rightarrow \textbf{Cat}$ assigning to each
context $ \Gamma $ the category whose objects are types in $\Ty_\C( \Gamma )$
and morphisms from $A$ to $B$ are substitutions
$ \varphi : \Gamma . A \rightarrow \Gamma . B$ such that
$\p \circ \varphi = \p $ -- those are in bijection with terms of type
$ \Gamma . A \vdash B[\p]$. The functorial action of $\TT$ is given by
\[
\TT( \gamma )( \varphi ) = \langle \p, \qI[ \varphi \circ (\gamma
\uparrow A)] \rangle : \Delta .A[\gamma]  \rightarrow \Delta .B[\gamma]
\]
for $\gamma : \Delta \to \Gamma$.

Any morphism $ \varphi $ in $\TT \Gamma $
from a type $A$ to a type $B$ induces a function
$\{ \varphi \} : \Tm_\C( \Gamma , A) \rightarrow \Tm_\C( \Gamma , B)$ which is
defined by
\[
\{ \varphi \}(a) = \qI[\varphi \circ \langle \id, a \rangle]
\] 

We will use this construction when transporting terms through
\emph{isomorphism of types} $\theta : A \cong_\Gamma B$, that is, 
isomorphisms in $\TT \Gamma $. We note the following:

\begin{lem}\label{lem:coerTT}
For any $\gamma : \Delta \to \Gamma$, $\varphi : \Gamma . A \to \Gamma . B$ in
$\TT \Gamma$, and $a \in \Tm_\C(\Delta, A[\gamma])$, 
\[
\{\TT( \gamma )( \varphi )\}(a) = \qI[ \varphi \circ \langle \gamma ,
a \rangle ]
\]
\end{lem}
\begin{proof}
Immediate from the definition.
\end{proof}

\begin{defi}[Pseudo cwf-morphisms]\label{def:pseudomor}
  A pseudo-cwf morphism from a cwf $\C$ to a cwf $\C'$ is a
  pair $(F, \sigma )$ where $F : \C \rightarrow \C'$ is a functor and
  for each $ \Gamma \in \Ctx_\C$, $ \sigma _ \Gamma $ is a
  \textbf{Fam}-morphism from $T \Gamma $ to $T'F \Gamma $ preserving
  the structure up to isomorphism. In particular there are isomorphisms
\[
\begin{array}{rrcl}
  \rho _{ \Gamma , A} :& F( \Gamma .A)  &\cong&  F \Gamma .FA\\
  \theta _{A,  \gamma } :& FA[F \gamma] &\cong_{F\Gamma}& F(A[\gamma])\hspace{30pt}\text{(for $\gamma :
\Gamma \to \Delta$)}\\
   !_{F} :& 1 &\cong& F1                                                          
\end{array}
\]
  satisfying some coherence diagrams, see Appendix \ref{app:pseudo_mor} for the complete definition.
\end{defi}

Since $ \sigma _ \Gamma $ is a \textbf{Fam}-morphism from
$(\Tm_\C( \Gamma , A))_{A  \in  \Ty_\C( \Gamma )}$ to $(\Tm_{\C'}(F \Gamma , B))_{B  \in  \Ty_{\C'}(F
\Gamma )}$ it has an action both on types and on terms.
We write $FA$ for the image of $A$ by the function $\Ty_\C( \Gamma )  \rightarrow  \Ty_{\C'}(F \Gamma )$
induced by $ \sigma _ \Gamma $, and $Fa$ for the image of $a\in \Tm_\C(\Gamma,A)$ through 
the function $\Tm_\C( \Gamma , A)  \rightarrow  \Tm_{\C'}(F \Gamma , FA)$ induced by $ \sigma _ \Gamma $.
As for cwfs, we will often refer to a pseudo cwf-morphism $(F, \sigma)$ just by $F$, keeping
$\sigma$ implicit. This goes in line with the notations introduced above, which do not mention
$\sigma$.

A pseudo cwf-morphism is strict whenever $ \theta _{A,  \gamma }$ and $ \rho _{ \Gamma ,A}$ are both
identities and $F1 = 1$. Cwfs and strict cwf-morphisms form a
category $\Cwfs$.

\begin{defi}[Pseudo cwf-transformation]\label{def:pseudo1}
  A pseudo cwf-transformation between pseudo cwf-morphisms $F$ and
  $G$ is a pair $( \varphi , \psi )$ where
  $ \varphi : F \Rightarrow G$ is a natural transformation, and for
  each $ \Gamma \in \Ctx_\C$ and $A \in \Ty_\C( \Gamma )$, 
  $ \psi _{ \Gamma , A}$ is a type isomorphism $FA  \cong_{F\Gamma}  GA [ \varphi _ \Gamma]$
  satisfying:
  $$ \varphi _{ \Gamma .A} = F( \Gamma .A) \xrightarrow{ \rho^F_{\Gamma, A}} F \Gamma .FA
\xrightarrow{ \psi _{ \Gamma , A}}F \Gamma .GA [\varphi _ \Gamma]  \xrightarrow{ \varphi _{ \Gamma
}^+} G \Gamma .GA \xrightarrow{ {\rho^G_{\Gamma, A}} ^{-1}} G( \Gamma .A),$$
\end{defi}

This means in particular that $\psi$ is uniquely determined from $\varphi$. However, it matches
our inductive proof later on to have both $\varphi$ and $\psi$ explicitely in the definition, with
this coherence diagram.
This definition corrects the one given in \cite{clairambault:mscs};
see Appendix \ref{app:pseudocwftrans} for a discussion on that.
We will write $\Cwf$ for the resulting 2-category.

\subsection{Syntax and inference rules for the free category with families}
\label{sec:syntax}

\subsubsection{Raw terms}\label{sec:grammar}
In this section we define the syntax and inference rules for a minimal dependent type theory with one
base type $o$. This theory is closely related to the generalised algebraic theory of cwfs \cite{dybjer:torino}, but here we define it as a usual logical system with a grammar and a collection of inference rules. The grammar has four syntactic categories: contexts $\RawCtx$,
substitutions $\RawSub$, types $\RawTy$ and terms $\RawTm$.
\begin{eqnarray*}
\Gamma \in \RawCtx &::=& 1  \ |\ \Gamma.A\\
\gamma \in \RawSub \ &::=& \gamma \circ \gamma \ |\ \id_\Gamma \ |\ \langle\rangle_\Gamma \ |\ \p_{A} \ |\ \langle \gamma, a \rangle_A\\
A \in \RawTy &::=& o \ |\  A[\gamma]\\
a \in \RawTm &::=& a[\gamma] \ |\ \qI_A
\end{eqnarray*}
These terms have as few annotations as possible, only what is
necessary to recover the domain and codomain of a substitution, the context of a type, and the type of a term:
\begin{align*}
\dom(\gamma \circ \gamma') &= \dom(\gamma') \quad&\quad\quad & \cod(\gamma \circ \gamma') &=&\  \cod(\gamma)\\
\dom(\id_\Gamma) &= \Gamma\quad& &\cod(\id_\Gamma) &=&\  \Gamma\\
\dom(\langle\rangle_\Gamma) &= \Gamma\quad&&\cod(\langle\rangle_\Gamma) &=&\  1\\
\dom(\p_{A}) &= \ctxof(A).A\quad&&\cod(\p_{A}) &=&\  \ctxof(A) \\
\dom(\langle \gamma, a \rangle_A) &= \dom(\gamma)\quad&&\cod(\langle \gamma, a \rangle_A) &=&\  \cod(\gamma).A\\
\ctxof(o) &= 1 \quad & & \typeof(a[\gamma]) &=&\ (\typeof(a))[\gamma]\\
\ctxof(A[\gamma]) &= \dom(\gamma) \quad& & \typeof(\qI_A) &=&\ A[\p_{A}]
\end{align*}
These functions will be used to define the interpretation.

\subsubsection{Inference rules}
We simultaneously inductively define four families of partial equivalence relations (pers) for the
four forms of equality judgments:
\[
\Gamma = \Gamma' \vdash \hspace{40pt} \Gamma \vdash A = A' \hspace{40pt}
\Delta \vdash \gamma = \gamma' : \Gamma \hspace{40pt} \Gamma \vdash a = a' : A
\]
In the inference rules which generate these pers we will use the following abbreviations for the basic judgment forms:
$\Gamma \vdash$ abbreviates $\Gamma = \Gamma \vdash$, 
$\Gamma \vdash A$ abbreviates $\Gamma \vdash A = A$,
$\Delta \vdash \gamma : \Gamma$ abbreviates  $\Delta \vdash \gamma = \gamma : \Gamma$, and 
$\Gamma \vdash a : A$ abbreviates $\Gamma \vdash a = a : A
$. The inference rules are divided into four kinds: \emph{per-rules}, which axiomatise symmetry and
transitivity of equality; \emph{preservation rules}, which express that equality preserves
judgments; \emph{congruence rules} for operators with respect to equality, and \emph{conversion
rules}.

\begin{figure}
  \centering
\boxit[Per-rules for the four forms of judgments]{
  \begin{mathpar}
    	\inferrule
		{\Gamma = \Gamma' \vdash \\ \Gamma' = \Gamma'' \vdash} 
		{\Gamma = \Gamma'' \vdash}
	\and 
	\inferrule
		{\Gamma = \Gamma' \vdash} 
		{\Gamma' = \Gamma \vdash}
	\and 
	\inferrule
		{\Delta \vdash \gamma = \gamma' : \Gamma \\ 
		 \Delta \vdash \gamma' = \gamma'' : \Gamma} 
		{\Delta \vdash \gamma = \gamma'' : \Gamma}
	\and 
	\inferrule
		{\Delta \vdash \gamma = \gamma' : \Gamma} 
		{\Delta \vdash \gamma' = \gamma : \Gamma}
	\and
   	\inferrule
		{\Gamma \vdash A = A' \\ 
		 \Gamma \vdash A' = A''} 
		{\Gamma \vdash A = A''}
	\and 
	\inferrule
		{\Gamma \vdash A = A'} 
		{\Gamma \vdash A' = A}
	\and 
	\inferrule
		{\Gamma \vdash a = a' : A \\ 
		 \Gamma \vdash a' = a'' : A} 
		{\Gamma \vdash a = a'' : A}
	\and 
	\inferrule
		{\Gamma \vdash a = a' : A}
		{\Gamma \vdash a' = a : A}
  \end{mathpar}}
  \label{fig:cwf1}
\end{figure}

\begin{figure}
  \centering
  \label{fig:cwf2}
\boxit[Preservation rules for judgments]{
  \begin{mathpar}
    	\inferrule
		{{\Gamma} = {\Gamma}' \vdashS \\
		 {\Delta} = {\Delta}' \vdashS \\
		 {\Gamma} \vdashS \gamma = \gamma': {\Delta}} 
		{{\Gamma}' \vdashS \gamma = \gamma' : {\Delta}'}
	\and
	\inferrule
		{{\Gamma} = {\Gamma}' \vdashS \\ 
		 {\Gamma} \vdashS A = A'}
		{{\Gamma}' \vdashS A = A'}
	\and 
	\inferrule
		{{\Gamma} = {\Gamma}' \vdashS \\ 
		 \Gamma \vdash A = A' \\
	         {\Gamma} \vdashS a = a' : A}
		{{\Gamma}' \vdashS a = a' : A'}
  \end{mathpar}
}
\end{figure}

\begin{figure}
  \centering
  \label{fig:cwf3}
\boxit[Congruence rules for operators and the base type]{
  \begin{mathpar}
    	\inferrule
		{ } 
		{1 = 1 \vdash }
	\and
	\inferrule
		{\Gamma = \Gamma' \vdash \\ 
		 \Gamma \vdash A = A'} 
		{\Gamma.A = \Gamma'.A'\vdash} 
	\and

	\inferrule
		{ } 
		{1 \vdash o = o} 
	\and
	\inferrule
		{\Gamma \vdash A=A' \\ 
		 \Delta \vdash \gamma = \gamma' : \Gamma} 
		{\Delta \vdash A[\gamma] = A'[\gamma']} 
	\and


	\inferrule
		{\Gamma = \Gamma' \vdashS } 
		{\Gamma \vdashS \id_\Gamma = \id_{\Gamma'} : \Gamma} 
	\and
    	\inferrule
		{\Gamma = \Gamma' \vdashS}
		{\Gamma \vdash \emptySub \Gamma = \emptySub {\Gamma'} : \emptyContext}
	\and
	\inferrule
		{{\Gamma} \vdashS \delta = \delta' : {\Delta} \\ 
		 {\Delta} \vdashS \gamma = \gamma': {\Theta} } 
		{{\Gamma} \vdashS \gamma \circ \delta = \gamma' \circ \delta' : {\Theta}} 
	\and
	\inferrule
    		{\Gamma \vdash A = A'}
		{\Gamma.A \vdash \p_{A} = \p_{A'} : \Gamma}
	\and
	\inferrule
		{\Gamma \vdash A = A' \\ 
		 \Delta \vdash \gamma = \gamma' : \Gamma \\ 
		 \Delta \vdash a = a' : A[\gamma]} 
		{\Delta\vdash \langle \gamma,a \rangle_A = \langle \gamma',a' \rangle_{A'} : \Gamma.A} 
	\and
	\inferrule
		{\Gamma \vdash a = a' : A \\
		 \Delta \vdash \gamma = \gamma' : \Gamma} 
		{\Delta \vdash a[\gamma] = a'[\gamma'] : A[\gamma]} 
	\and 

	\inferrule
		{\Gamma \vdash A = A'}
		{\Gamma.A \vdash \qI_{A} = \qI_{A'} : A[\p_A]} 
    
  \end{mathpar}
}
\end{figure}

\begin{figure}
  \centering
  \label{fig:cwf4}
\boxit[Conversion rules]{
  \begin{mathpar}
     	\inferrule
		{\Delta \vdash \theta : \Theta \\ 
		 \Gamma \vdash \delta : \Delta \\ 
		 \Xi \vdash \gamma : \Gamma} 
		{\Xi \vdash (\theta \circ \delta) \circ \gamma = \theta \circ (\delta \circ \gamma) : \Theta} 
	\and
    	\inferrule
		{\Gamma \vdashS \gamma : \Delta} 
		{\Gamma \vdash \gamma = \id_\Delta \circ \gamma : \Delta}
	\and 
	\inferrule
		{\Gamma \vdashS \gamma : \Delta} 
		{\Gamma \vdash \gamma = \gamma \circ \id_\Gamma : \Delta}
	\and 
	\inferrule
		{\Gamma \vdash A \\ 
		 \Delta \vdash \gamma : \Gamma \\ 
		 \Theta \vdash \delta : \Delta} 
		{\Theta \vdash A[\gamma \circ \delta] = (A[\gamma])[\delta]}
	\and
	\inferrule
		{\Gamma \vdash A}
		{\Gamma \vdash A[\id_\Gamma] = A}
	\and
	\inferrule
		{\Gamma \vdash a : A \\ 
		 \Delta \vdash \gamma : \Gamma \\ 
		 \Theta \vdash \delta : \Delta} 
		{\Theta \vdash a[\gamma \circ \delta] = (a[\gamma])[\delta] : (A[\gamma])[\delta]}
	\and
    	\inferrule
		{\Gamma \vdash a : A}
		{\Gamma \vdash a[\id_\Gamma] = a : A} 
	\and 
	\inferrule
		{\Gamma \vdash \gamma : 1} 
		{\Gamma \vdash \gamma = \emptySub{\Gamma} : 1} 
	\and 
	\inferrule
		{\Gamma \vdash A \\ 
		 \Delta \vdash \gamma : \Gamma \\ 
		 \Delta \vdash a : A[\gamma]}
		{\Delta \vdash \p_A \circ \langle \gamma,a \rangle_A = \gamma : \Gamma} 
	\and
	\inferrule
		{\Gamma \vdash A \\ 
		 \Delta \vdash \gamma : \Gamma \\
		 \Delta \vdash a : A[\gamma]} 
		{\Delta \vdash \qI_A[\langle \gamma,a \rangle_A] = a : A[\gamma]} 
	\and 
	\inferrule
		{\Delta \vdash \gamma : \Gamma.A}
		{\Delta \vdash \gamma = \langle \p_A \circ \gamma , \qI_A[\gamma] \rangle_A : \Gamma.A}
  \end{mathpar}
}
\end{figure}

Note that our syntax is annotated in order to ensure that a raw term has a unique (up to judgmental
equality) type given by the function $\typeof$, and that a type has a unique (up to judgemental equality) context given by the function
$\ctxof$. Similarly, $\domain$ and $\codomain$ return the unique domain and codomain of a substitution.

\begin{lem}\label{lemma:unique_typing}
  We have the following:
  \begin{itemize}
  \item If $ \Gamma   \vdash  A$ is derivable, then $  \Gamma  = \ctxof (A)\vdash  $ is also derivable.
  \item If $ \Gamma   \vdash  a : A$ is derivable, then $ \Gamma  = \ctxof (A)\vdash   $ and $
\Gamma   \vdash A = \typeof (a)$ are derivable.
  \item If $ \Delta   \vdash   \gamma  :  \Gamma $ is derivable, then $  \Delta  = \domain(
\gamma )\vdash  $ and $ \Gamma  = \codomain( \gamma )\vdash   $ are derivable.
  \end{itemize}
\end{lem}

\subsubsection{The syntactic cwf $\T$}
We can now define a term model as the syntactic cwf obtained by the well-formed contexts, substitutions, types, and terms, modulo judgmental equality. We use brackets for equivalence classes in this definition. (Note that brackets are also used for substitution in types and terms. However, this should not cause confusion since we will soon drop the equivalence class brackets.)

\begin{defi}
The term model $\T$ is given by:
\begin{itemize}
\item $\Ctx_\T = \{ {\Gamma}\ |\ \Gamma \vdashS \} /\!\! =^c$, where
  ${\Gamma} =^c {\Gamma}'$ if ${\Gamma} = {\Gamma}' \vdashS$ is
  derivable.
\item
  $\Sub_\T([{\Gamma}],[{\Delta}]) = \{ \gamma\ |\ \Gamma \vdashS \gamma
  : {\Delta} \} /\!\! =^{\Gamma}_{\Delta}$
  where $\gamma =^{\Gamma}_{\Delta} \gamma'$ iff
  ${\Gamma} \vdashS \gamma = \gamma' : {\Delta}$ is derivable. Note that
  this makes sense since it only depends on the equivalence class of
  $\Gamma$ (morphisms and morphism equality are preserved by object
  equality).
\item $\Ty_\T([{\Gamma}]) = \{ A\ |\ \Gamma \vdashS A
  \}/=^{\Gamma}$ where $A =^{\Gamma} B$ if $\Gamma \vdashS A =
  B$.
\item $\Tm_\T([{\Gamma}],[A]) = \{ a\ |\ \Gamma \vdashS a: A\} / =^{\Gamma}_A$ where $a =^{\Gamma}_A
a'$ if $\Gamma \vdashS a = a' : A$. 
\end{itemize}
The cwf-operations on $\T$ can now be defined in a straightforward way. For example, if $\Delta \vdash \theta : \Theta$, $\Gamma \vdash \delta : \Delta$, we define
$
[\theta ] \circ_\T [\delta] = [\theta \circ \delta],
$
which is well-defined since composition preserves equality.
\end{defi}
\subsection{Freeness of $\T$}
\label{sec:freenessT}

We shall show that $\T$ is a free cwf on one base type, in the sense that for an arbitrary cwf $\C$ and type $o_\C  \in  \Ty_\C(1_\C)$, there exists
a unique strict cwf morphism $\T  \rightarrow  \C$ which maps $ [o]$
to $o_\C$. Such a morphism can be defined by first defining a partial
function for each sort of raw terms (where $\RawCtx$ denotes the set
of raw contexts, $\RawSub$ the set of raw substitutions, and so on
defined by the grammar of Section \ref{sec:grammar}), cf Streicher
\cite{streicher:book}. 
\begin{eqnarray*}
\intCtx{-} &:& \RawCtx\ \rightharpoonup \Ctx_\C\\
\intCtx{-}&:& (\gamma \in \RawSub)\ \rightharpoonup \
\Sub_\C(\domain(\gamma), \codomain(\gamma))\\
\intCtx{-}&:& (A \in \RawTy)\ \rightharpoonup\ \Ty_\C(\ctxof(A))\\
\intCtx{-}&:& (t \in\RawTm)\ \rightharpoonup\
\Tm_\C (\ctxof(\typeof(t)), \typeof(t))
\end{eqnarray*}

We use the notation 
$(x \in A) \rightharpoonup B(x)$ for the {\em partial} dependent function space, that is, the set of partial functions $f$ which map $x \in A$ to $f(x) \in B(x)$ whenever $f(x)$ is defined.

Note that we use the same notation for all four interpretation functions.
These partial interpretation functions are defined by mutual induction on the structure of raw terms:

\[
\begin{array}{rclcrclcrcl}
\inte{1} &=& 1_\C 						&&
\inte{\gamma' \circ \gamma} &= & \inte{\gamma'} \circ_\C \inte{\gamma}&&
\inte{ \langle  \rangle_ \Gamma  } &=&  (\langle  \rangle_\C) _{ \inte \Gamma  }
\\
\intCtx { \Gamma .A} &=& \ \inte \Gamma ._\C \inte{A}	&&
\inte{\id_{\Gamma}} & = &\ {(\idC)}_{\inte{\Gamma}}	&&
\inte {a[\gamma]} &=& \inte a[\inte \gamma]_\C
\\
\inte {o} &=& o_\C					&&
\inte{ \langle  \gamma , a \rangle_A } &=& \langle \inte \gamma , \inte a \rangle_\C&&
\inte {\qI_A} &= & (\qI_\C)_{\inte A}
\\
\inte{A[\gamma]} &=& \inte A [\inte \gamma]_\C		&&
\inte {\p_A} &= & (\p_\C)_{\inte A} 
\end{array}
\]

%
Partiality arises because, for instance, $\inte{\gamma'} \circ_\C {\inte{\gamma}}$ is only defined when $\inte{\gamma'}$ and $\inte{\gamma}$ are defined and $\dom(\inte{\gamma'}) = \cod(\inte{\gamma})$.
However, we can prove by induction on the inference rules that the interpretation of equal well-formed contexts, equal well-typed substitutions, equal well-formed types, and equal well-typed terms are always defined and equal:
\begin{lem}\hfill
\label{lemma:interp_defined}
  \begin{itemize}
  \item If $\Gamma = \Gamma' \vdash$, then
    both $\intCtx{\Gamma}$ and $\intCtx{\Gamma'}$ are defined in $\Ctx_\C$, and equal.
  \item If $\Delta \vdash \gamma = \gamma' : \Gamma$, then
    $\inte{\gamma} = \inte{\gamma'} \in \Sub_\C(\intCtx{\Delta},{\intCtx{\Gamma}})$ are defined and
equal.
  \item If $\Gamma \vdash A = A'$, then
    $\inte{A} = \inte{A'} \in \Ty_\C(\intCtx{\Gamma})$ are defined and equal.
  \item If $\Gamma \vdash a = a' : A$, then
    $\inte{a} = \inte{a'} \in \Tm_\C(\intCtx{\Gamma},\inte{A})$ are defined and equal.
  \end{itemize}
\end{lem}

It follows in particular that if we have $\Gamma \vdash$ (which abbreviates $\Gamma = \Gamma \vdash$), then $\intCtx{\Gamma}$ is defined -- and likewise for the other
reflexive judgements.
Hence, we can define total interpretation functions on the term model by restricting the partial 
interpretation function to the well-formed contexts, etc, and then lift it to the quotient:
\begin{eqnarray*}
\iniCtx{-} &:& \Ctx_\T\ \arrow\ \Ctx_\C\\
\iniSub{[\Gamma]}{[\Delta]}{-} &:&
 \Sub_\T([\Gamma],{[\Delta]}) \ \arrow \
 \Sub_\C(\iniCtx{[\Gamma]},{\iniCtx{[\Delta]}})\\
\iniTy{[\Gamma]}{-} &:& \Ty_\T([\Gamma])\ \arrow\ \Ty_\C(\iniCtx{[\Gamma]})\\
\iniTm{[\Gamma]}{[A]}{-} &:& \Tm_\T([\Gamma],[A])\ \arrow\ \Tm_\C(\iniCtx{[\Gamma]},\iniTy{[\Gamma]}{[A]})
\end{eqnarray*}
by
\begin{eqnarray*}
\iniCtx{[\Gamma]} &=& \inte{\Gamma}\\
\iniSub{[\Gamma]}{[\Delta]}{[\gamma]} &=& \inte{\gamma}\\
\iniTy{[\Gamma]}{[A]} &=& \inte{A}\\
\iniTm{[\Gamma]}{[A]}{[a]} &=& \inte{a}
\end{eqnarray*}
which is well-defined by Lemma \ref{lemma:unique_typing}.

This defines a strict cwf
morphism $\T  \rightarrow  \C$ which maps 
$[o]$ to $o_\C$. 
In order to prove that it is unique, we assume that $F : \T \arrow \C$ is another strict cwf
morphism, and prove by induction on the inference rules (the pers) that if $\Gamma = \Gamma' \vdash$
then $F[\Gamma] = \iniCtx{[\Gamma]}$, etc. For example, $1 = 1 \vdash$ and we prove $F[1] = 1_\C =
\iniCtx{[1]}$ by preservation of the terminal object. The other cases are similarly straightforward,
since strict cwf-morphisms preserve the structure on the nose.

This concludes the proof of our theorem:
\begin{thm}\label{theom:initial_cwf}
  $\T$ is a free cwf on one object, that is, for every other cwf $\C$
  and $o_\C  \in  \Ty_\C(1_\C)$ there is a unique strict cwf morphism $\T  \rightarrow  \C$
  which maps $[o]$ to $o_\C$.
\end{thm}

It is in fact \emph{the} free cwf on one object up to isomorphism, since any two free cwfs are related by a unique
isomorphism.

From now on we will uniformly drop the equivalence class brackets and for example write $\Gamma$ for $[\Gamma]$. There should be no risk of confusion, but we remark that proofs by induction on syntax and inference rules are on representatives rather than equivalence classes.

\subsection{Bifreeness of $\T$}\label{a-free-cwf}
We eventually wish to add the type formers $\N_1, \Sigma, \Pi$ and $\I$, and construct the free cwf which supports these type formers. 
However, as we explained in the introduction, this freeness property will not transport to lcccs. Indeed, our
correspondence between cwfs (with support for these type formers) and lcccs is a \emph{biequivalence}
\cite{clairambault:mscs} rather than an equivalence, and freeness is not preserved by biequivalence
(the dimensions of the two notions mismatch -- likewise, isomorphism is
not preserved by biequivalence, but \emph{equivalence} is).
Moreover, so far we proved that $\T$ is free in the category of cwfs and \emph{strict} cwf-morphisms which 
preserve cwf-structure on the nose. In lcccs, finite limits and local exponents are usually not treated as
extra \emph{structure}, but as \emph{properties} of categories. Thus functors can only
preserve these properties up to \emph{isomorphism}, since it would not even make sense to say that these properties are
preserved on the nose. As a consequence, in our biequivalence result
we moved to \emph{pseudo} cwf-morphisms (Definition \ref{def:pseudomor}) that only preserve structure up to
coherent isomorphism. The cwf $\T$ is not free in the category of cwfs with pseudo cwf-morphisms --
in fact, there is no free cwf in this category. However, we can move to a 2-categorical setting and show that $\T$ is \emph{bifree}.

We recall that an object $I$ is bi-initial in a 2-category iff for any
object $A$ there exists an arrow $I \to A$ and for any two arrows
$f, g : I \to A$ there exists a unique 2-cell
$\theta : f \Rightarrow g$. It follows that $\theta$ is invertible, and that 
bi-initial objects are equivalent. 

\begin{defi} \label{def:cwf_o}A cwf $\C$ is {\em bifree} on one base type iff it is
  bi-initial in the 2-category $\Cwf^o$:
  \begin{itemize}
  \item \emph{Objects}: pairs $(\C, o_\C)$ where $\C$ is a cwf and $o_\C  \in  \Ty_\C(1_\C)$.
  \item \emph{Morphisms between $(\C, o_\C)$ and $(\D, o_\D)$}: pairs
    $(F,\alpha_F)$ of pseudo cwf-morphisms $F : \C \rightarrow \D$ and
    isomorphisms $ \alpha_F: F(o_\C)[ !_F ] \cong o_\D$ in the category of
    closed types $\textbf T(1_\D)$ (recall that $!_F : 1  \rightarrow  F1$).
      \item \emph{2-cells between the morphisms $(F, \alpha_F), (G,\alpha_G) : (\C, o_\C)  \rightarrow  (\D, o_\D)$}: pseudo cwf-trans\-for\-mations $( \varphi ,  \psi )$
    from $F$ to $G$ satisfying
    $ \psi _{o_\C} =  \alpha _G^{-1} \circ  \alpha _F : F(o_\C)[!_F]  \cong_{1_\D}  G(o_\C)[!_G].$
  \end{itemize}
\end{defi}

\noindent The rest of the section is dedicated to the proof of the following:
\begin{thm}
  $\T$ is a bifree cwf on one base type.
\end{thm}
We have shown that for every cwf $\C$, and $o_\C \in \Ty_\C(1_\C)$, the
interpretation $\iniCtx{-}$ is a strict cwf-morphism mapping $o$ to
$o_\C$. Hence it is a morphism in $\Cwf^o$. It remains to show that for
any other morphism $F : \T \rightarrow \C$ in $\Cwf^o$, there is a
unique 2-cell (pseudo cwf-transformation)
$( \varphi , \psi ) : \iniCtx{-} \to F$, which is an isomorphism.
This asymmetric version of bi-initiality is equivalent to
that given above.

\subsubsection{Existence of $( \varphi ,  \psi )$} \label{sec:ex_cwf}
We construct $( \varphi ,  \psi )$ by induction on the inference rules and simultaneously prove their naturality properties:
\begin{itemize}
\item If $ \Gamma = \Gamma'  \vdash $, then there exists \emph{an isomorphism} $ \varphi _ \Gamma = \varphi _ {\Gamma'}  : \interp  \Gamma   \cong  F\Gamma $.
\item If $ \Gamma \vdash A = A'$, then there exists \emph{an isomorphism}
  $ \psi _A = \psi _{A'}: \inte {A}  \cong_{\interp  \Gamma } FA[\varphi _ \Gamma]$.
\item If $ \Gamma   \vdash   \gamma = \gamma' :  \Delta $, then $F\gamma \circ \varphi _ \Gamma= \varphi _ \Delta \circ \interp \gamma$.
\item If $ \Gamma   \vdash  a = a' : A$, then $Fa [\varphi _ \Gamma] = \{ \psi _A\}(\interp a)$.
\end{itemize}\enlargethispage{\baselineskip}\smallskip

It follows that $( \varphi , \psi )$ is a pseudo cwf-transformation.
We show some crucial cases:\smallskip

\noindent\textbf{Empty context.} $F$ preserves terminal objects and we let $\varphi_1 =\ !_F: \interp { 1 } = 1_\C \cong F 1$.

\medskip

\noindent\textbf{Context extension.} By induction, we have
  $ \psi _A : \interp A  \cong FA [\varphi _ \Gamma]$ in $\TT( \Gamma )$.
  We define $ \varphi _{ \Gamma .A}$ as the following composition of isomorphisms:
  $$ \interp{\Gamma . A} = \interp{\Gamma}.\interp{A} \xrightarrow{ \psi _A} \interp \Gamma .FA
[\varphi _ \Gamma] \xrightarrow{ \langle  \varphi _ \Gamma \circ \p, \qI \rangle } F \Gamma .FA \xrightarrow {  \rho _{ \Gamma ,A}^{-1} } F( \Gamma .A)$$
We remark that this case of the induction concerns the rule that not only expresses the well-formedness of context extension, but more generally, that context extension preserves equality. So officially, we need to prove that 
$ \varphi _ \Gamma = \varphi _ {\Gamma'}  : \interp  \Gamma   \cong  F\Gamma $ and $ \psi _A = \psi _{A'}: \inte {A}  \cong_{\interp  \Gamma } FA[\varphi _ \Gamma]$ entail $ \varphi _{ \Gamma .A} = \varphi _{ \Gamma' .A'}$ which follows immediately. We also remark that we have dropped the official index $A$ in $\p_A$ and $\qI_A$ in the above definition. Both remarks apply in other cases too.\smallskip

\noindent\textbf{Base type.} By definition, $F$
  is equipped with $ \alpha _F
  :  \interp o  \cong  F(o)[!_F]$. We define $\psi_o =  \alpha _F :  \interp o  \cong 
  F(o)[!_F]$ in $\Ty_\C(1)$.

\medskip

\noindent\textbf{Type substitution.} Let $ \Gamma \vdash  \gamma  : \Delta $
  and $ \Delta \vdash A$. The induction hypotheses are
  $ \varphi _ \Delta  \circ \interp  \gamma  = (F  \gamma)  \circ \varphi _ \Gamma$ and
  $ \psi _A : \interp A \cong _ {\interp \Delta} FA [\varphi _ \Delta]$. Since
  $\TT$ is a contravariant functor, $\TT  \interp{\gamma} $ is a functor from $\TT  \interp{\Delta} $ to $\TT
\interp{\Gamma} $ thus,
  $$\TT(\interp  \gamma )( \psi _A) : \interp {A[\gamma] }  \cong _{\interp \Gamma}  FA [\varphi _
\Delta \circ  \interp\gamma] = FA [F \gamma] [\varphi _ \Gamma]$$
  by induction hypothesis on $ \gamma $. So we define
  $$ \psi _{A[\gamma] } = \TT( \varphi _ \Gamma )( \theta _{A,  \gamma }) \circ \TT(\interp  \gamma )( \psi _A)  : \interp {A[\gamma] }  \cong_{\interp \Gamma}  (F(A[\gamma] )) [\varphi _ \Gamma] $$

\medskip

\noindent\textbf{Projection.} We have $\Gamma . A  \vdash \p_A : \Gamma $ and need to check
  that $F\p_A \circ \varphi _{ \Gamma .A}= \varphi_\Gamma \circ \p$. This is a simple calculation:
  \begin{align*}
     F\p_A \circ \varphi _{ \Gamma .A} &=  F\p \circ \rho _{ \Gamma ,A}^{-1} \circ \langle  \varphi _
\Gamma \circ \p, \qI \rangle \circ \psi _A \quad&&\text{(definition of $ \varphi _{ \Gamma .A}$)} \\
    &=  \p \circ \langle \varphi _ \Gamma \circ \p, \qI \rangle \circ \psi _A\quad&&\text{(property of $ \rho _{ \Gamma , A}$)} \\
    &=  \varphi _ \Gamma \circ \p \circ \psi _A = \varphi _ \Gamma \circ \p \quad&&\text{(because $
\psi _A$ is a map in $\TT  \interp{\Gamma} $)}
  \end{align*}

\medskip

\noindent\textbf{Extension.} Assume we have $ \Gamma   \vdash  \gamma :  \Delta $ and $ \Gamma   \vdash  t : A[\gamma]$ so
  that $ \langle \gamma, t \rangle_{A} $ is a morphism from $ \Gamma $ to $ \Delta   .  A$. 
  Using Proposition \ref{prop:pres_subst_ext}, we get that $F  \langle \gamma, t \rangle_{A}  \circ  \varphi _ \Gamma = \rho ^{-1}_{\Delta, A} \circ  \langle
F\gamma, \{
\theta _{A,\gamma}^{-1}\} (Ft) \rangle  \circ  \varphi _ \Gamma$. After calculation, we get
$$
\begin{array}{rcll}
  F  \langle \gamma, t \rangle_A  \circ  \varphi _ \Gamma &=& \rho ^{-1}_{\Delta, A} \circ  \langle
F\gamma, \{
\theta _{A,\gamma}^{-1}\} (Ft) \rangle  \circ  \varphi _ \Gamma \\
  &= &  \rho ^{-1}_{\Delta, A} \circ  \langle F\gamma \circ  \varphi _ \Gamma , \{ \theta
_{A,\gamma}^{-1}\}(Ft) [\varphi _ \Gamma] \rangle \\
& &\text{(Lemma \ref{lem:comp_subst_coer})}\\
  &= &  \rho ^{-1}_{\Delta, A} \circ  \langle F\gamma \circ  \varphi _ \Gamma , \{
\TT(\varphi_\Gamma)(\theta_{A,\gamma}^{-1})\}(Ft [\varphi_\Gamma]) \rangle \\
& &\text{(I.H. on $\gamma$ and $t$)} \\
  &= &  \rho ^{-1}_{\Delta, A} \circ  \langle  \varphi _ \Delta  \circ \interp{\gamma}, \{\TT(\varphi _
\Gamma)( \theta
_{A,\gamma}^{-1})\}( \{\psi _{A[\gamma]}\}(\interp{t})) \rangle \\ 
& & \text{(definition of $ \psi _{A[\gamma]}$)} \\
  &= &  \rho ^{-1}_{\Delta,A} \circ  \langle  \varphi _ \Delta  \circ \interp{\gamma},
\{\TT(\interp{\gamma})( \psi
_A)\}(\interp{t})
\rangle \\
& &\text{(definition of $\TT$)} \\
  &= &  \rho ^{-1}_{\Delta,A} \circ  \langle  \varphi _ \Delta  \circ \interp{\gamma}, \qI [\psi _A \circ
\langle \interp{\gamma}, \interp{t}
\rangle] \rangle\\
  &= &  \rho ^{-1}_{\Delta,A} \circ  \langle  \varphi _ \Delta  \circ \p, \qI \rangle  \circ  \psi _A
\circ  \interp{\langle
\gamma, t \rangle_A}  \\
  &=&  \rho ^{-1}_{\Delta,A} \circ  \varphi _ \Delta   ^+  \circ  \psi _A \circ \interp{ \langle \gamma, t
\rangle_A}  \\
  &=& \varphi _{ \Delta .A} \circ  \interp{\langle \gamma, t \rangle_A}
\end{array}
$$

\noindent\textbf{Term substitution.} Assume we have $ \Gamma \vdash \gamma : \Delta $
  and $ \Delta \vdash t : A$. Unfolding the definition of $\psi_{A[\gamma]}$, we get:
\[
\begin{array}{rcll}
\{\psi_{A[\gamma]}\}(\interp{t[\gamma]}) &=&
\{\TT(\varphi_\Gamma)(\theta_{A, \gamma})\}(\{\TT(\interp{\gamma})(\psi_A)\}(\interp{t[\gamma]})) \\
&=& \{\TT(\varphi_\Gamma)(\theta_{A,
\gamma})\}(\{\TT(\interp{\gamma})(\psi_A)\}(\interp{t}[\interp{\gamma}]))\\
& &\text{(Lemma \ref{lem:comp_subst_coer})}\\
&=& \{\TT(\varphi_\Gamma)(\theta_{A, \gamma})\}(\{\psi_A\}(\interp{t})[\interp{\gamma}])
\\ 
& & \text{(I.H. on $t$)}\\
&=& \{\TT(\varphi_\Gamma)(\theta_{A, \gamma})\}((Ft)[\varphi_\Delta \circ \interp{\gamma}])\\
 & &\text{(I.H. on $\gamma$)}\\
&=& \{\TT(\varphi_\Gamma)(\theta_{A, \gamma})\}((Ft)[F\gamma \circ \varphi_\Gamma]) \\
& &\text{(Lemma \ref{lem:comp_subst_coer})}\\
&=& \{\theta_{A, \gamma}\}(Ft [F\gamma])[\varphi_\Gamma] \\ 
& &\text{(Definition of pseudo cwf-morphisms)} \\
&=& F(t[\gamma])[\varphi_\Gamma]\\
\end{array}
\]

\medskip

\noindent\textbf{Variable.} Assume we have $ \Gamma . A \vdash \qI_A :
  A [\p]$. Unfolding the definition of $ \psi _{A[\p]}$ yields, after some
simplifications:
\begin{align*}
\{ \psi _{A[\p]}\}(\qI_A) &= \qI [\theta _{A, \p} \circ  \langle  \varphi _{ \Gamma }, \qI [\psi _A \circ  \langle \p, \qI \rangle ] \rangle ] \\
&\qquad\text{($ \langle \p, \qI \rangle  = \id$)}\\
&=\qI [\theta _{A, \p} \circ  \langle  \varphi _{ \Gamma  . A}, \qI [\psi _A] \rangle ]
\end{align*}
We need to prove that this is equal to:
\begin{align*}
F\qI [\varphi _{ \Gamma  . A}] &= \{ \theta _{A, \p}\} \left(\qI\left[\rho_{\Gamma, A}\right]\right)[\varphi _{ \Gamma .A}] \\
&\qquad\text{(definition of pseudo cwf-morphism)} \\
&= \qI [\theta _{A, \p} \circ  \langle \id, \qI [\rho_{\Gamma, A}]  \rangle  \circ  \varphi _{ \Gamma .A}] \\
&=\qI [\theta _{A, \p} \circ  \langle  \varphi _{ \Gamma .A}, \qI [\rho_{\Gamma,A}  \circ  \varphi _{ \Gamma  . A}] \rangle ] \\
&\qquad \text{(definition of $ \varphi _{ \Gamma .A}$)} \\
&= \qI [\theta _{A, \p} \circ  \langle  \varphi _{ \Gamma .A}, \qI [\langle  \varphi _ \Gamma \circ
\p , \qI \rangle  \circ  \psi _A \rangle ] \\
&=\qI [\theta _{A, \p} \circ  \langle  \varphi _{ \Gamma .A}, \qI [\psi _A] \rangle ]
\end{align*}
Thus the equality holds.

\medskip

\noindent\textbf{Functoriality of substitution.}
Assume we have $ \Gamma   \vdash  \gamma :  \Delta $ and
  $ \Delta   \vdash  \delta :  \Theta $. We want to show the equality $ \psi _{A[\delta][\gamma]} =  \psi _{A[\delta\circ \gamma]}$
  and $ \psi _{A[\id]} =  \psi _A$ for $ \Theta   \vdash  A$. The second equation is easy: by
  functoriality of $\TT$, $\TT(\interp {\id})( \psi _A) =  \psi _A$ and
  properties of $F$, $ \theta _{A, \id} = \id$. \enlargethispage{\baselineskip}

  For the other equation, unfolding the definitions gives:
\begin{align*}
   \psi _{A[\delta][\gamma]} &= \TT (\interp{\gamma}) (\TT (\interp{\delta})( \psi _A) \circ \TT( \varphi _ \Delta )( \theta _{A, \delta})) \circ \TT ( \varphi _ \Gamma )( \theta _{A, \gamma}) \\
&= \TT(\interp \gamma) \left( \TT(\interp{\delta})( \psi _A) \right) \circ \TT(\interp{\gamma}) \left(\TT( \varphi _ \Delta )( \theta _{A, \delta}\right)) \circ \TT( \varphi _ \Gamma )( \theta _{A, \gamma}) \\
&\qquad\text{(functoriality of $\TT$ and induction hypothesis on $\gamma$)} \\
&= \TT(\interp \delta \circ \interp \gamma)( \psi _A) \circ \TT( \varphi _ \Gamma )(\TT(F\gamma)( \theta _{A, \delta})) \circ \TT( \varphi _ \Gamma )( \theta _{A, \gamma})\\
&\qquad\text{(functoriality of $\TT( \varphi _ \Gamma )$)} \\
&= \TT(\interp \delta \circ \interp \gamma)( \psi _A) \circ \TT( \varphi _ \Gamma )(\TT(F\gamma)( \theta _{A, \delta}) \circ ( \theta _{A, \gamma})) \\
&\qquad\text{(coherence for $ \theta $)} \\
 &= \TT(\interp \delta \circ \interp \gamma)( \psi _A) \circ \TT( \varphi _ \Gamma )( \theta _{A, \delta \circ \gamma})\\
&=  \psi_ {A[\delta \circ \gamma]}
\end{align*}
Other cases arising from conversion rules and per-rules are straightfoward.

\subsubsection{Uniqueness of $(
  \varphi , \psi )$}\label{sec:cwf_uniq} Let $( \varphi ', \psi '):
\interp \cdot \to F$ be another pseudo cwf-transformation in
$\Cwf^o$. We prove the following by induction:

\begin{itemize}
\item If $ \Gamma =  \Gamma '   \vdash $, then $ \varphi _ \Gamma  =  \varphi '_ \Gamma $
\item If $ \Gamma   \vdash  A = A'$, then $ \psi _A =  \psi '_A$
\end{itemize}

\medskip

\noindent\textbf{Empty context.} There is a unique morphism between the
  terminal objects $\interp {1}$ and $F1$, so $ \varphi _{1} =  \varphi '_{1}$.

\medskip

\noindent\textbf{Context extension.} Assume by induction
  $ \varphi _ \Gamma = \varphi _ \Gamma '$ and $ \psi _A = \psi _A'$.
  By the coherence law of pseudo cwf-transformations, we have
$ \varphi '_{ \Gamma .A} =  \rho^{-1}_{\Gamma, A} \circ {\varphi '_ \Gamma}  ^+ \circ \psi '_A    $ from which the
  equality $\varphi _{ \Gamma .A} = \varphi '_{ \Gamma .A} $ follows.

\medskip

\noindent\textbf{Type substitution.} Assume we have $ \Delta \vdash A$ and
  $ \Gamma \vdash \gamma : \Delta $, and consider
  $ \psi _{A[\gamma] }$ and $ \psi '_{A[\gamma] }$. By Lemma \ref{lem:transcoh}, we have:
\[
\psi_{A[\gamma]}' = \TT( \varphi '_ \Gamma )( \theta _{A, \gamma }) \circ
  \TT(\interp{\gamma} )( \psi '_A)
\]
  and likewise for $\psi_{A[\gamma]}$.  Since we know by induction
  hypothesis that $ \varphi _ \Gamma = \varphi _ \Gamma '$ and
  $\psi_{A} = \psi'_A$, it follows that
  $\psi_{A[\gamma]} = \psi_{A[\gamma]}'$.

\medskip

\noindent\textbf{Base type.} The definition of 2-cells in $\Cwf^o$ 
entails $ \psi '_{o} = \alpha_F^{-1} :  \interp o  \rightarrow  F( \llbracket o \rrbracket)$.

This concludes the proof that $\T$ is a bifree cwf on one object. In the next section, we will prove
that this result still holds in the presence of type constructors.

\section{A free lccc}\label{a-free-lcc}


This section will basically follow the plan of Section \ref{sec:freecwf}. We will first recall what it
means for categories with families to support the extra structure for $\I$, $\N_1$, $\Pi$ and
$\Sigma$-types. Then we will extend our cwf-calculus with these type constructors. Finally, we will
also extend our proofs of freeness and bifreeness. In particular, bifreeness will be transported by our biequivalence
\cite{clairambault:mscs}. It follows that the underlying category of contexts of the syntactic cwf with extra structure is a 
bifree lccc.

\subsection{Cwfs which support $\I, \N_1, \Sigma, \Pi$}

We recall here from \cite{dybjer:torino,clairambault:mscs} what it
means for a cwf to support type constructors and prove a few properties of the
corresponding combinators.


\begin{defi}
A cwf $\C$ supports extensional identity types iff it is equipped with the following extra structure:
\begin{itemize}
\item \emph{Formation.} If $A \in \Ty_\C(\Gamma)$ and $a, a' \in \Tm_\C(\Gamma, A)$, then there is
$\I(A, a, a') \in \Ty_\C(\Gamma)$.
\item \emph{Introduction.} If $a \in \Tm_\C(\Gamma, A)$, then there is 
$\refl(a) \in \Tm_\C(\Gamma,\I(A, a, a))$.
\item \emph{Elimination.} If $c \in \Tm_\C(\Gamma, \I(A, a, a'))$, then $a = a'$ and $c = \refl(a)$.
\end{itemize}
such that the following laws with respect to substitution are satisfied, for any $\gamma : \Delta
\to \Gamma$:
\begin{eqnarray*}
\I(A, a, a')[\gamma] &=& \I(A[\gamma], a[\gamma], a'[\gamma])\\
\refl(a)[\gamma] &=& \refl{(a[\gamma])}
\end{eqnarray*}
\end{defi}

\begin{defi}
A cwf $\C$ supports $\Sigma$-types iff it is equipped with the following extra structure:
\begin{itemize}
\item \emph{Formation.} If $A\in \Ty_\C(\Gamma)$ and $B\in \Ty_\C(\Gamma . A)$, there is $\Sigma(A,B)\in
\Ty_\C(\Gamma)$,
\item \emph{Introduction.} If $a \in \Tm_\C(\Gamma,A)$ and $b \in \Tm_\C(\Gamma,B[\ext{\id,a}])$, there is
$\pair(a,b) \in \Tm_\C(\Gamma, \Sigma(A,B))$,
\item \emph{Elimination.} If $c \in \Tm_\C(\Gamma, \Sigma(A,B))$, there are $\fst( c ) \in
\Tm_\C(\Gamma, A)$ and $\snd( c ) \in \Tm_\C(\Gamma, B[\ext{\id,\fst( c )}])$ such that
\begin{eqnarray*}
\fst(\pair(a,b)) &=& a\\
\snd(\pair(a,b)) &=& b\\
\pair(\fst(c),\snd(c)) &=& c
\end{eqnarray*}
\end{itemize}
and we also have stability under substitution. If $\gamma : \Delta \to \Gamma$ then
\begin{eqnarray*}
\Sigma(A,B)[\gamma] &=& \Sigma(A[\gamma],B[\ext{\gamma\circ \p,\qI}])\\
\pair(a,b)[\gamma] &=& \pair(a[\gamma],b[\gamma])\\
\fst(c)[\gamma] &=& \fst(c[\gamma])\\
\snd(c)[\gamma] &=& \snd(c[\gamma])
\end{eqnarray*}
\end{defi}

Before going on to the definition of cwfs supporting $\Pi$-types, it is useful to recall a few
lemmas about $\Sigma$-types on cwfs. First we recall from \cite{clairambault:mscs}:

\begin{lem}\label{lem:chi}
For any $A \in \Ty_\C(\Gamma)$ and $B \in \Ty_\C(\Gamma.A)$, there is an isomorphism:
\[
\chi_{A, B} : \Gamma . A . B \to \Gamma . \Sigma(A, B)
\]
such that $\p \circ \chi_{A, B} = \p \circ \p$.
\end{lem}
\begin{proof}
The isomorphism is defined by the following inverse substitutions:
\begin{eqnarray*}
\ext{\p\circ\p, \pair(\qI[\p], \qI)} &:& \Gamma . A . B \to \Gamma . \Sigma(A, B)\\
\ext{\ext{\p, \fst(\qI)}, \snd(\qI)} &:& \Gamma . \Sigma(A, B) \to \Gamma . A . B
\end{eqnarray*}
An easy calculation shows that they are mutual inverses.
\end{proof}

The type constructor $\Sigma$ can also be extended to act on morphisms in the adequate fibres,
in a functorial way. This is formalized in the following lemma.

\begin{lem}\label{lemma:sigmafunctorial}
Let $A, A' \in \Ty_\C(\Gamma), B \in \Ty_\C(\Gamma . A)$, and $B' \in \Ty_\C(\Gamma . A')$.
Moreover, consider morphisms $f_A : A \to A'$ in $\TT(\Gamma)$ (\emph{i.e.} $f_A : \Gamma . A \to \Gamma
. A'$ such that $\p \circ f_A = \p$), and $f_B : B \to B'[f_A]$ in $\TT(\Gamma . A)$ (\emph{i.e.} $f_B :
\Gamma . A . B \to \Gamma . A . B'[f_A]$ such that $\p \circ f_B = \p$).

Then, defining:
\[
\Sigma(f_A, f_B) : \ext{\p,\pair(\qI[f_A \circ \ext{\p, \fst(\qI)}], \qI[f_B\circ \ext{\ext{\p, \fst(\qI)},
\snd(\qI)}])}
\]
we have $\Sigma(f_A, f_B) : \Sigma(A, B) \to \Sigma(A', B')$ in $\TT(\Gamma)$. Moreover, it is
functorial in the following sense. For $f_A, f_B$ as above and $g_A : A' \to A'', 
g_B : B' \to B''[g_A]$, then:
\[
\Sigma(g_A, g_B) \circ \Sigma(f_A, f_B) = \Sigma(g_A \circ f_A, \TT(f_A)(g_B) \circ f_B)
\]
\end{lem}

\begin{proof}
Direct verification.
\end{proof}

This strengthens Lemma B.1 of \cite{clairambault:mscs}, which states that the type constructor
$\Sigma$ preserves isomorphisms of types. We will also use the following lemma, which states
compatibility of the functorial action of $\Sigma$ with that of substitution. 

\begin{lem}\label{lem:sig_funct_subst}
Let $f_A, f_B$ be as in the lemma above. Then, for any $\gamma : \Delta \to \Gamma$, we have:
\[
\TT( \gamma )(\Sigma(f_A, f_B)) =  \Sigma (\TT( \gamma )(f_A), \TT( \gamma \uparrow A)(f_B))
\]
Both are morphisms from $\Sigma(A[\gamma], B[\gamma \uparrow A])$ to $\Sigma(A'[\gamma], B'[\gamma
\uparrow A])$ 
in $\TT( \Delta )$.
\end{lem}
\begin{proof}
Direct calculation.
\end{proof}

Now, we go on to define what it means for a cwf to support $\Pi$-types.

\begin{defi}
A cwf $\C$ supports $\Pi$-types iff it is equipped with the following extra structure:
\begin{itemize}
\item \emph{Formation.} If $A\in \Ty_\C(\Gamma)$ and $B\in \Ty_\C(\Gamma . A)$, there is $\Pi(A,B)\in
\Ty_\C(\Gamma)$.
\item \emph{Introduction.} If $b \in \Tm_\C({\Gamma}.{A},B)$, there is $\lambda(b) \in
\Tm_\C(\Gamma,\Pi (A,B))$.
\item \emph{Elimination.} If $c \in \Tm_\C(\Gamma,\Pi(A,B))$ and $a \in \Tm_\C(\Gamma, A)$ then there is a
term $\ap(c,a) \in \Tm_\C(\Gamma, B[\ext{\id,a}])$ such that
\begin{eqnarray*}
\ap(\lambda (b),a) &=& b[\ext{\id,a}]\\
\lambda (\ap(c[\p],\qI )) &=& c 
\end{eqnarray*}
\end{itemize}
and we also have stability under substitution. If $\gamma : \Delta \to \Gamma$ then
\begin{eqnarray*}
{\Pi (A,B)}[\gamma] &=&
\Pi ({A}[{\gamma}],{B}[\ext{\gamma \circ \p,\qI}])\\
(\lambda (b))[\gamma] &=& \lambda (b[\ext{\gamma \circ \p,\qI}])\\
(\ap(c,a))[\gamma] &=& \ap(c[\gamma],a[\gamma])
\end{eqnarray*}
\end{defi}

Just like for $\Sigma$-types, $\Pi$-types can be given a functorial action on the fibres.

\begin{lem}\label{lemma:pifunctorial}
Let $A, A' \in \Ty_\C(\Gamma), B \in \Ty_\C(\Gamma . A)$, and $B' \in \Ty_\C(\Gamma . A')$.
Moreover, consider morphisms $f_A : A' \to A$ in $\TT(\Gamma)$ and $f_B : B[f_A] \to B'$ in $\TT(\Gamma . A')$.

Then, defining:
\[
\Pi(f_A, f_B) = \ext{\p, \lambda(\qI[f_B \circ \ext{\ext{\p \circ \p, \qI}, \ap(\qI[\p],\qI[f_A
\circ \ext{\p \circ \p, \qI}])}])}
\]
we have $\Pi(f_A, f_B) : \Pi(A, B) \to \Pi(A', B')$ in $\TT(\Gamma)$. Morever, the action of $\Pi$
is functorial, in the sense that for $f_A, f_B$ as above and $g_A : A'' \to A'$, $g_B : B'[g_A] \to
B''$, we have:
\[
\Pi(g_A, g_B) \circ \Pi(f_A, f_B) = \Pi(f_A \circ g_A, g_B \circ \TT(g_A)(f_B))
\]
\end{lem}
\begin{proof}
Tedious calculations on cwf-combinators.
\end{proof}

Just as for $\Sigma$-types, the functorial action of $\Pi$ commutes with the functorial action of
substitution.

\begin{lem}\label{lemma:piT}
Let $f_A, f_B$ as in the lemma above, and $\gamma : \Delta \to \Gamma$. Then, we have:
\[
\TT( \gamma )(\Pi(f_A, f_B)) =  \Pi (\TT( \gamma )(f_A), \TT( \gamma  \uparrow A')(f_B))
\]
where both terms are morphisms from $\Pi (A[ \gamma ], B[ \gamma  \uparrow A])$ to 
$\Pi (A'[ \gamma ], B'[ \gamma  \uparrow A'])$
in $\TT( \Delta )$.
\end{lem}
\begin{proof}
Direct calculation.
\end{proof}


\begin{defi}
A cwf $\C$ supports $\N_1$ iff it is equipped with the following extra structure:
\begin{itemize}
\item \emph{Formation.} There is $\N_1 \in \Ty_\C(1)$.
\item \emph{Introduction.} There is $0_1 \in \Tm_\C(1, \N_1)$.
\item \emph{Elimination.} For any $c \in \Tm_\C(1, \N_1)$, $c = 0_1$.
\end{itemize}
\end{defi}

We will be interested in cwfs that support $\N_1$. However, both the cwfs that come from syntax
(including $\T$) and the cwfs in correspondence with lcccs through our biequivalence satisfy a
stronger property: they are \emph{democratic}.

\begin{defi}[Democratic cwfs]
  A cwf $\C$ is democratic when for each context $ \Gamma $ there is a type
  $\bar  \Gamma   \in  \Ty_\C(1)$ with an isomorphism $ \gamma_\Gamma : \Gamma   \cong  1.\bar  \Gamma $.
\end{defi}


\begin{lem}\label{lem:unit}
Let $\C$ be a democratic cwf. Then, it supports $\N_1$. 
\end{lem}
\begin{proof}
We simply define $\N_1 = \bar 1$. This type automatically has an inhabitant $0_1 =
\qI[\gamma_1] \in \Tm_\C(1, \bar 1)$; its uniqueness is an easy consequence of the fact that $1$ is
terminal.
\end{proof}

As a consequence we do not need to mention support for $\N_1$ for democratic cwfs . We will show in Lemma \ref{lem:syndem} that in the presence of $\Sigma$-types and
$\N_1$, the syntactically generated cwf is democratic.

For each of these type constructors, it is easy to define what it means for strict cwf-morphisms to
preserve them. We simply ask that everything -- both type constructors and the associated
combinators -- is preserved on the nose. For instance, we ask that 
\[
F(\Gamma. \Sigma(A, B)) = F\Gamma . \Sigma(FA, FB)
\]
and $F(\pair(a, b)) = \pair(F a, F b)$, \emph{etc.}. 

However, as emphasized before, for the correspondence with lcccs one needs notions of cwf-morphisms
that only preserve structure \emph{up to isomorphism.}

\subsection{Pseudo cwf-morphisms preserving structure up to isomorphism}

We now recall the definitions of preservation of structure up to
isomorphism for pseudo cwf-morphisms from \cite{clairambault:mscs}.
Note first that for  cwfs which support $\Sigma$-types, pseudo cwf-morphisms automatically preserve $\Sigma$-types.

\begin{prop}\label{prop:preservsigma}
A pseudo cwf-morphism $F$ from $\C$ to $\C'$, where both cwfs support
$\Sigma$-types, also preserves them
in the sense that there is an isomorphism:
\[
s_{A, B} : F(\Sigma(A, B)) \cong \Sigma(FA, FB[\rho_{\Gamma, A}^{-1}])
\]
such that projections are preserved up to isomorphism. For any term $c \in \Tm_\C(\Gamma,\Sigma(A,
B))$, or terms $a\in \Tm_\C(\Gamma,A)$ and $b\in \Tm_\C(\Gamma,B[\ext{\id, a}])$.
\begin{eqnarray*}
F(\fst(c)) &=& \fst(\{s_{A, B}\}(Fc))\\
F(\snd(c)) &=& \{\theta_{B, \ext{\id,\fst(c)}}\}(\snd(\{s_{A, B}\}(Fc)))\\
F(\pair(a, b)) &=& \{s_{A, B}^{-1}\}(\pair(F a,
\{\theta_{B, \ext{\id, a}}^{-1}\}(F b)))
\end{eqnarray*}
\end{prop}
\begin{proof}
   Proposition 3.5 in \cite{clairambault:mscs}.
\end{proof}

On the other hand, neither the preservation of the other type constructors nor the preservation of democracy is automatic. We recall the following definitions from \cite{clairambault:mscs}. 

\begin{defi}
Let $\C, \C'$ be cwfs supporting identity types and $F : \C \to \C'$ be a pseudo cwf-morphism. Then, 
$F$ preserves identity types provided there is an isomorphism:
\[
F(\Id(A,a,a')) \cong \Id(FA,Fa,Fa')
\]
in $\TT'(\Gamma)$.
\end{defi}

Likewise, we have for democracy:

\begin{defi}
Let $\C, \C'$ be democratic cwfs, and $F : \C \to \C'$ be pseudo cwf-morphisms. Then, $F$ preserves
democracy provided there is an isomorphism
\[
d_\Gamma : F(\bar{\Gamma}) \cong \bar{F\Gamma}[\ext{}]
\]
in $\TT'(1)$ such that the following diagram commutes:
\[
\xymatrix@R=10pt{
F\Gamma \ar[rr]^{F\gamma_\Gamma}
        \ar[d]_{\gamma_{F\Gamma}}&&
F(1 . \bar{\Gamma})
        \ar[d]^{\rho_{1, \bar{\Gamma}}}\\
1 . \bar{F\Gamma}
        \ar@{<-}[r]^{\ext{\ext{}, \qI}}&
F1 . \bar{F\Gamma}[\ext{}]
        \ar@{<-}[r]^{d_\Gamma}&
F1 .  F(\bar{\Gamma})
}
\]
\end{defi}

We saw before that democratic cwfs automatically support $\N_1$ -- likewise, pseudo cwf-morphisms
that preserve democracy automatically preserve $\N_1$ in the obvious sense. 

Finally, we define preservation of $\Pi$-types.

\begin{defi}
Let $\C$ and $\C'$ be cwfs supporting $\Pi$-types, and $F$ a pseudo
cwf-morphism. Then $F$ preserves $\Pi$-types
iff for each types $A\in \Ty_\C(\Gamma)$ and $B \in \Ty_\C(\Gamma . A)$ there is an isomorphism in
$\TT'(\Gamma)$:
\[
i_{A, B} : F(\Pi(A, B)) \cong \Pi(F(A), F(B)[\rho_{\Gamma, A}^{-1}])
\]
such that for any substitution $\gamma : \Delta \to \Gamma$, for any terms $c\in \Tm_\C(\Delta,\Pi(A,
B)[\gamma])$ and $a \in \Tm_\C(\Gamma,A[\gamma])$, we have:
\[
F(\ap(c, a)) = 
\{\theta_{B, \ext{\gamma, a}}\}(\ap(\{\TT'(F\gamma)(i_{A, B}) \circ \theta_{\Pi(A, B),
\gamma}^{-1}\}(Fc), \{\theta_{A, \gamma}^{-1}\}(Fa)))
\]
\end{defi}

The definition of preservation of $\Pi$-types for pseudo cwf-morphisms only
require them to preserve application. In fact, as remarked in \cite{clairambault:mscs}, it is
sufficient to ensure that abstraction is preserved as well.

\begin{lem}\label{lemma:pilambda}
If $F : \C \to \C'$ is a pseudo cwf-morphism preserving $\Pi$-types, then it preserves the
abstraction combinator, in the sense that for any $b \in \Tm_\C(\Gamma . A, B)$, 
\[
F(\lambda(b)) = \{i_{A, B}^{-1}\}(\lambda((Fb)[\rho_{\Gamma, A}^{-1}]))
\]
\end{lem}
\begin{proof}
Immediate consequence of Lemma \ref{lem:carac_pres_pi}.
\end{proof}

We now go on to extend our syntactic cwf $\T$ with all the extra structure mentioned above, before proving
that it is bifree.

\subsection{The syntactic cwf with extensional $\I$, $\N_1$, $ \Sigma $, and $ \Pi$}

We extend the grammar and the set of inference rules with rules for $\I, \N_1, \Sigma,$ and $\Pi$-types:
\begin{eqnarray*}
A &::=& \cdots \ |\   \I(A,a,a)  \ |\  {\N_1} \ |\ \Sigma(A,A)  \ |\  \Pi(A,A)\\
a &::=& \cdots \ |\   \refl(a) \ |\  {0_1} \ |\ \fst(A,a)  |\  \snd(A,A,a)  |\  \pair(A,A,a,a)  |\  \ap(A,A,a,a) |\  \lambda(A,a)
\end{eqnarray*}
For each type we define its context:
\begin{align*}
\ctxof(\I(A,a,a')) &= \ctxof(A)\\
\ctxof(\N_1) &= 1 \\
\ctxof(\Sigma(A,B)) &= \ctxof(A) \\
\ctxof(\Pi(A,B)) &= \ctxof(A)
\end{align*}
%

For each term we define its type:
\begin{align*}
\typeof(0_1) &= \N_1\\
\typeof(\fst(A,c)) &= A\\
\typeof(\snd(A,B,c) &= B\,[\langle \id_{\ctxof(A)},\fst(A,c)\rangle_{A}]\hspace{-20pt}\\
\typeof(\pair(A,B,a,b)) &= \Sigma(A,B) \\
\typeof(\refl(a)) &= \I(\typeof(a),a,a) \\
\typeof(\lambda(A,c)) &= \Pi(A,\typeof(c))\\
\typeof(\ap(A,B,c,a)) &= B\,[\langle \id_{\ctxof(A)},a \rangle_{A}]
\end{align*}

There is still some redundancy in the type annotations: one could omit the
annotation $A$ in $\I(A, a, a')$ and only have $\I(a, a')$. Its context can then be
recovered as $\ctxof(\typeof(a))$ instead of $\ctxof(A)$. However, this optimization makes the
termination of the mutually recursive functions $\ctxof$ and $\typeof$ less immediately evident as it is no
longer a simple structural induction\footnote{We are grateful to one of the reviewers for this
observation.}. We therefore opted for the present slightly redundant version.

\begin{figure}[H]
\boxit[Rules for $\I$-types]{
\begin{mathpar}
    	\inferrule
		{ \Gamma \vdash a = a': A \and \Gamma \vdash b = b' : A} 
		{ \Gamma \vdash \I(A,a,b) = \I(A,a',b')} 
	\and 
	\inferrule 
		{ \Gamma \vdash a = a': A}
		{ \Gamma \vdash \refl(a) = \refl(a') : \I(A,a, a')} 
	\and 
	\inferrule 
		{ \Gamma \vdash c : \I(A,a, a')} 
		{ \Gamma \vdash a = a' : A} 
	\and 
	\inferrule 
		{ \Gamma \vdash c : \I(A,a,a')} 
		{ \Gamma \vdash c = \refl(a) : \I(A,a,a')} 
	\and 
	\inferrule
		{ \Gamma \vdash a : A \\ \Gamma \vdash a' : A \\ \Delta \vdash \gamma : \Gamma}
		{ \Delta \vdash \I(A,a, a')[\gamma] = \I(A[\gamma],a[\gamma] , a'[\gamma])} 
  \end{mathpar}
}
\end{figure}
\begin{figure}[H]
  \centering
\boxit[Rules for $\N_1$]{
  \begin{mathpar}
	\inferrule
		{ }
		{ 1 \vdash \N_1 = \N_1 }
\and
	\inferrule
		{ }
		{ 1 \vdash 0_1 = 0_1 : \N_1 }
\and
	\inferrule
		{ 1 \vdash a : \N_1 }
		{ 1 \vdash a = 0_1 : \N_1 }
  \end{mathpar}
}
\end{figure}

\begin{figure}
  \label{fig:sigma}
\boxit[Rules for $\Sigma$-types]{
\begin{mathpar}
\inferrule 
	{\Gamma \vdash A = A' \\ 
	 \Gamma.A \vdash B = B'}
	{\Gamma \vdash \Sigma(A,B) = \Sigma(A',B')} 
\and
\inferrule 
	{\Gamma \vdash A = A' \\ 
	 \Gamma \vdash c = c' : \Sigma(A,B)} 
	{\Gamma \vdash \fst(A,c) = \fst(A',c') : A} 
\and 
\inferrule 
	{\Gamma \vdash A = A' \\
	 \Gamma.A \vdash B = B' \\
	 \Gamma \vdash c = c' : \Sigma(A,B)}
	{\Gamma \vdash \snd(A,B,c) = \snd(A',B',c') : B\,[\langle \id_{\Gamma}, \fst(A,c)\rangle_{A}] } 
\and
\inferrule
	{\Gamma \vdash A = A' \\ 
	 \Gamma.A \vdash B = B' \\
	 \Gamma \vdash a = a' : A' \\
	 \Gamma \vdash b = b' : B\,[\langle \id_{\Gamma}, \fst(A,c)\rangle_{A}]}
	{\Gamma \vdash \pair(A,B,a,b) = \pair(A',B',a',b') : \Sigma(A,B)} 
\and
\inferrule
	{\Gamma \vdash A \\ 
	 \Gamma.A \vdash B \\ 
	 \Gamma \vdash a : A \\
	 \Gamma \vdash b : B\,[\langle \id_{\Gamma}, \fst(A,c)\rangle_{A}]} 
	{\Gamma : \fst(A,\pair(A,B,a,b)) = a : A} 
\and 
\inferrule
	{\Gamma \vdash A \\
	 \Gamma.A \vdash B \\ 
	 \Gamma \vdash a : A \\
	 \Gamma \vdash b : B\,[\langle \id_{\Gamma}, \fst(A,c)\rangle_{A}]}
	{\Gamma \vdash \snd(A,B,\pair(A,B,a,b)) = b : B\,[\langle \id_{\Gamma},\fst(A,c)\rangle_{A}]} 
\and
\inferrule
    	{\Gamma \vdash c : \Sigma(A, B)}
	{\Gamma \vdash c = \pair(A,B,\fst(A,c),\snd(A,B,c)) :  \Sigma (A, B)}
\and
\inferrule
	{\Gamma \vdash A \\ 
	 \Gamma.A \vdash B \\ 
	 \Delta \vdash \gamma : \Gamma}
	{\Delta \vdash \Sigma(A,B)[\gamma] = \Sigma(A[\gamma],B[\gamma^+])} 
\and
\inferrule
	{\Gamma \vdash A \\ 
	 \Gamma \vdash c : \Sigma(A,B) \\ 
	 \Delta \vdash \gamma : \Gamma} 
	{\Delta \vdash \fst(A,c)[\gamma] = \fst(A[\gamma],c[\gamma]) : A} 
\and 
\inferrule
	{\Gamma \vdash A \\ 
	 \Gamma.A \vdash B \\ 
	 \Gamma \vdash c : \Sigma(A,B) \\ 
	 \Delta \vdash \gamma : \Gamma} 
    	{\Delta \vdash \snd(A,B,c)[\gamma] = \snd(A[\gamma],B[\gamma^+], c[\gamma]) : B[\langle  \gamma , \fst(A, c)[\gamma]  \rangle_{A}] } 
\and
\inferrule 
	{\Gamma \vdash A \\ 
	 \Gamma.A \vdash B \\ 
	 \Gamma \vdash a : A \\ 
	 \Gamma \vdash b : B[\langle \id_{\Gamma}, \fst(A,c)\rangle_{A}] \\ 
	 \Delta \vdash \gamma : \Gamma}
	{\Delta \vdash \pair(A,B,a,b)[\gamma] =\pair(A[\gamma],B[\gamma^+], a[\gamma], b[\gamma]) :  \Sigma (A, B)[\gamma] } 
\end{mathpar}}
\end{figure}

\begin{figure}[H]
\boxit[Rules for $\Pi$-types]{
\begin{mathpar}
\inferrule 
	{\Gamma \vdash A = A' \\
	 \Gamma.A \vdash B = B'}
	{\Gamma \vdash \Pi(A,B) = \Pi(A',B')} 
\and
\inferrule
	{ \Gamma   \vdash  A = A' \\  
	 \Gamma .A  \vdash  b = b' : B} 
	{ \Gamma  \vdash   \lambda (A, b) =  \lambda (A', b') :  \Pi (A, B)}
\and
\inferrule
	{ \Gamma   \vdash  A = A' \\  
	 \Gamma .A  \vdash  B = B' \\  
	 \Gamma   \vdash  c = c' :  \Pi (A, B) \\  
	 \Gamma   \vdash  a = a' : A} 
	{ \Gamma   \vdash  \app(A, B, c, a) = \app(A', B', c', a') : B[\langle \id_{\Gamma}, a \rangle_{A}] }
\and
\inferrule
	{ \Gamma.A \vdash b : B \\  
	 \Gamma   \vdash  a : A}
	{ \Gamma   \vdash  \app(A, B,  \lambda (A, b), a) = b[\langle \id_{\Gamma}, a \rangle_{A}]  : B [\langle \id_{\Gamma}, a \rangle_{A}] }
\and
\inferrule
	{ \Gamma   \vdash  c :  \Pi (A, B)}
	{ \Gamma   \vdash   \lambda (A, \app(c[p], q)) = c :  \Pi (A, B)}
\and
\inferrule
	{ \Gamma   \vdash  A \\  
	 \Gamma .A  \vdash  B \\  
	 \Delta   \vdash   \gamma  :  \Gamma } 
	{ \Delta   \vdash   \Pi (A, B) [\gamma]  =  \Pi (A [\gamma] , A [\gamma ^+])}
\and
\inferrule
	{ \Gamma   \vdash  c :  \Pi (A, B) \\ 
	 \Delta  \vdash   \gamma  :  \Gamma } 
	{ \Delta   \vdash   \lambda (A,b)[\gamma]  =  \lambda (A[\gamma],b [\gamma ^+]) :  \Pi (A, B) [\gamma] }
\and
\inferrule
	{ \Gamma   \vdash  c :  \Pi (A, B) \\  
	 \Gamma   \vdash  a : A \\ 
	 \Delta  \vdash   \gamma  :  \Gamma }
	{ \Delta   \vdash  \app(c, a) [\gamma]  = \app(c [\gamma] , a [\gamma] ) : B [\langle \gamma , a [\gamma]  \rangle_{A}] }
  \end{mathpar}}
  \label{fig:pi}
\end{figure}

It is straightforward to extend the definition of the term model $\T$
with $\I, \N_1, \Sigma$, and $\Pi $-types to form a cwf
$\T^{\I, \N_1, \Sigma , \Pi}$ supporting these type constructors. Although there are no grammatical construct and no inference rules corresponding to democracy we can prove the following:

\begin{lem}\label{lem:syndem}
The cwf $\T^{\I, \N_1, \Sigma , \Pi}$ is democratic.
\end{lem}
\begin{proof}
For any well-formed context $\Gamma \vdash$ we define a type $\bar \Gamma$ by induction on the inference rules. For $1 \vdash$, we have $\bar{1} = \N_1 \in \Ty(1)$. For $\Gamma . A \vdash$, we set
$\bar{\Gamma . A} = \Sigma(\bar \Gamma, A[\gamma_\Gamma^{-1}])$. Constructing the required isomorphism is immediate by
induction using Lemma \ref{lem:chi}.
\end{proof}

It is straightforward to extend the interpretation functor and prove its uniqueness (among
\emph{strict} cwf-morphisms). It is also easy to check that it preserves democracy.

\begin{thm}
  $\T^{\I, \N_1, \Sigma , \Pi}$ is the free democratic cwf supporting
  $\I, \Sigma , \Pi $ on one object.
\end{thm}

We do not detail the proof of this theorem: in essence, it is a simplified version of the proof of
Theorem \ref{thm:bifreecwfextra} where all key isomorphisms are replaced by identities. Instead,
we go on to prove that just as $\T$, besides being free in the category of strict cwf-morphisms
(preserving structure), $\T^{\I, \N_1,\Sigma ,  \Pi }$ is also bifree in the $2$-category of
\emph{pseudo} cwf-morphisms (preserving structure).

\subsection{Bifreeness of $\T^{\I, \N_1,\Sigma ,  \Pi }$}

We now prove the key result:
\begin{thm}\label{thm:bifreecwfextra}
  $\T^{\I, \N_1, \Sigma , \Pi}$ is the bifree democratic cwf supporting
  $\I, \Sigma , \Pi $ on one object.
\end{thm}
This means that $\T^{\I, \N_1, \Sigma , \Pi}$ is bi-initial in the 2-category
  $\Cwf_d^{\I, \Sigma , \Pi , o}$ where objects are democratic cwfs which support $\I, \Sigma ,
\Pi$, and a base type $o$,  and where morphisms preserve these type formers up to coherent
isomorphisms.

\subsubsection{Existence of $\varphi$ and $\psi$}

We resume our inductive proof from Section \ref{sec:ex_cwf}, treating the additional inference rules
for $\I, \N_1, \Sigma$ and $\Pi$. We will first treat the type formation rules, then the type
substitution rules. The rules for conversion and substitution on terms are straightforward, and not
detailed.

\medskip

\noindent\textbf{Type formation rules.} We start with the type formation rules for $\N_1, \I, \Sigma$
and $\Pi$.
\begin{description}
\item[Unit type] Since $F$ preserves democracy and the terminal object it follows that:
$$1.\bar 1  \cong  1  \cong  F1  \cong  1.\overline{F1}  \cong  1.F(\N_1)[ \varphi _1]$$

Write $\psi_{\bar{1}}$ for this type isomorphism.
\item[Identity type] Assume $ \Gamma   \vdash  a, b, a', b': A$. By induction hypothesis, we have
  $ \psi _A : \interp A  \cong_{\interp  \Gamma}  FA[ \varphi _ \Gamma ]$. We know $\I$-types preserve isomorphisms in the
  indexed category (Lemma \ref{lem:prop_id_iso}) yielding (over $\interp \Gamma$):
  \begin{align*}
    \psi _{\I(A,a, b)}: \interp{\I(A,a, b)} 
	&= \I(\interp{A},\interp{a}, \interp{b})\\
	&\cong \I(FA[
    \varphi _ \Gamma ], \{ \psi _A\}(\interp a), \{ \psi _A\}(\interp
    {b}))\\ &= \I(FA[ \varphi _ \Gamma ], F(a)[ \varphi _ \Gamma ],
    F(b)[ \varphi _ \Gamma ])
  \end{align*}

We also have $\psi_{\I(A,a', b')}$ defined likewise. But $\psi_{\I(A,a, b)}$ and
$\psi_{\I(A,a',
b')}$ are two parallel type isomorphisms whose domain is an identity type -- so
$\qI[\psi_{\I(A,a, b)}]^{-1}, \qI[\psi_{\I(A,a', b')}^{-1}] \in
\Tm_\C(\interp{\Gamma}.F(\I(A,a,
b))[\varphi_\Gamma], \I(\interp{A}[\p], \interp{a}[\p], \interp{b}[\p])$. It follows by 
the elimination rule for identity types in a cwf that these are both equal to the reflexivity term, and
that $\psi_{\I(A,a, b)} = \psi_{\I(A,a', b')}$.

\item[$\Sigma$-types] Assume that we have $\Gamma \vdash A = A'$
and $\Gamma.A \vdash B = B'$. By induction we have the isomorphisms
$ \psi _A = \psi_{A'} : \interp A  \cong _{\interp{\Gamma}}  FA[ \varphi _ \Gamma ]$ and
$ \psi _B = \psi_{B'} : \interp B  \cong _{\interp{\Gamma .A}} FB[ \varphi _{ \Gamma .A}]$.
We let:
  \begin{align*}
\psi _{ \Sigma (A, B)} =\  \interp{\Gamma . \Sigma (A, B)} &\xrightarrow{ \Sigma ( \psi _A, \psi
_B)} \interp{\Gamma}
. \Sigma (FA[ \varphi  _ \Gamma ], FB[ \rho_{\Gamma, A}^{-1} \circ {\varphi_ \Gamma}   ^+])
 \\& \xrightarrow{ \TT( \varphi _ \Gamma )(s _{A, B}^{-1})} \interp{\Gamma}
  .F( \Sigma (A, B))[ \varphi _ \Gamma ]
  \end{align*}

It is clear by construction that $\psi_{\Sigma(A, B)} = \psi_{\Sigma(A', B')}$. 
\item[$\Pi$-types] Consider $\Gamma \vdash A = A'$ and $\Gamma.A \vdash B = B'$.
 Define $ \psi _{ \Pi (A, B)}$ as follows:
  \begin{align*}
    \interp{ \Gamma . \Pi (A, B)} \xrightarrow{ \Pi ( \psi _A^{-1},  \TT( \psi _A^{-1})(\psi _B))}\
& \interp{ \Gamma} . \Pi (FA[ \varphi _ \Gamma ], FB[ \rho_{\Gamma, A}^{-1} \circ  \varphi _{
\Gamma}^+]) \\
    \xrightarrow{\TT( \varphi _ \Gamma )( i _{A, B}^{-1})}\ &{ \interp \Gamma .F( \Pi (A, B))[ \varphi _ \Gamma ]}
  \end{align*}

It is clear by construction that $\psi_{\Pi(A, B)} = \psi_{\Pi(A', B')}$.
\end{description}

\medskip

\noindent\textbf{Type substitution rules.} We now deal with the inference rules pertaining to the
compatibility of the types $\I, \Sigma$ and $\Pi$ with substitution. There is no inference rule for
compatibility of $\N_1$ with substitution.

In order to deal with compatibility under substitution, it is convenient to start with a few lemmas.
In particular, we will use heavily the fact that $\theta_{A, \gamma}$ can be characterised with a
universal property.

\begin{lem}\label{lem:theta_lemma}
  Let $\gamma :  \Gamma   \rightarrow   \Delta $. The type morphism $ \theta _{A, \gamma}$ is the
only type
  morphism to make the following diagram commute:
$$\xymatrix {
  F \Gamma .F(A[\gamma]) \ar[r]^{ \rho_{ \Gamma , A[\gamma]}^{-1} } & F(  \Gamma  .A[\gamma])
\ar[r]^{F (\gamma^+)} & F(  \Delta  .A) \ar[d]_{ \rho _{  \Delta  , A}} \\
F  \Gamma  .FA[F\gamma]\ar[u]_{ \theta _{A, \gamma}} \ar[rr]_{ (F\gamma)^+ } & & F \Delta . FA
}$$
\end{lem}
\proof
  The diagram commutes by virtue of Lemma \ref{lem:pres_lift_sub}. Moreover, by definition of type
  substitution the following diagram is a pullback:
  $$\xymatrix {
    \ar[d]_\p F \Gamma .FA[F\gamma] \pb{315}\ar[rr]^{(F\gamma)^+} & & F \Delta .FA \ar[d]^\p \\
    F \Gamma  \ar[rr]_{F \gamma } & & F \Delta
  }$$
  Because $ \theta $ is an isomorphism and the diagram above commutes, the following is also a
pullback:
  $$\xymatrix@C=40pt{
    \ar[d]_\p    F \Gamma .F(A[\gamma]) \pb{315} \ar[rr]^{ \rho_{\Delta, A}  \circ F(\gamma^+)
\circ \rho_{\Gamma, A[\gamma]} ^{-1}} & & F
\Delta .FA \ar[d]^\p \\
    F \Gamma \ar[rr]_{F\gamma} & & F \Delta }$$
  Thus it follows that there is a unique type morphism $F \Delta .FA[F\gamma]  \rightarrow  F \Delta
.F(A[\gamma])$ that makes the
  diagram of the lemma commute by the universal property of pullbacks.
\qed

Using that, we deduce two lemmas on the compatibility of $\Sigma$-types and $\Pi$-types under
substitution.

\begin{lem}[Compatibility of $ \Sigma $-types with substitution]\label{lemma:sigmaTh}
For any $A\in \Ty_\C(\Delta), B \in \Ty_\C(\Delta.A)$ and $\gamma : \Gamma \to \Delta$,
the following diagram of type isomorphisms over $F \Gamma $ commutes.
$$\xymatrix@C=3.5cm {
\ar[d]_{\TT(F\gamma)(s _{A, B})} F( \Sigma (A, B))[F\gamma] \ar[r]^{ \theta _{ \Sigma (A, B),\gamma}} & F(
\Sigma (A, B)[\gamma]) \ar[d]^{{ s _{A[\gamma], B[\gamma^+]}}} \\
 \Sigma (FA, FB[ \rho ^{-1}_{\Delta,A}])[F\gamma] \ar[r]_{ \Sigma ( \theta _{A,\gamma},
\TT(\rho_{\Gamma, A[\gamma]}^{-1}\circ \theta_{A, \gamma})( \theta
_{B,\gamma^+}))} &  \Sigma (F(A[\gamma]), F(B[\gamma^+])[\rho_{\Gamma, A[\gamma]}^{-1}])
}$$
(It is well-typed because of the diagram of Lemma \ref{lem:theta_lemma})
\end{lem}
\proof
  The diagram amounts to showing that
  $ \theta _{ \Sigma (A, B), \gamma} =  s^{-1} _{A[\gamma], B[\gamma^+]} \circ  \Sigma ( \theta
_{A,\gamma},
\TT(\rho_{\Gamma, A[\gamma]}^{-1}\circ \theta_{A, \gamma})( \theta
_{B,\gamma^+}))
  \circ T(F\gamma)( s _{A, B})$.
  Hence by Lemma \ref{lem:theta_lemma} it is enough to show that the right hand
  side makes the corresponding diagram commute -- which is an involved
  calculation.
\qed

\begin{lem}[Compatibility of $ \Pi $-types with substitution]
  \label{lemma:piTh}
For any $A\in \Ty_\C(\Delta), B \in \Ty_\C(\Delta.A)$ and $\gamma : \Gamma \to \Delta$,
the following diagram of type isomorphisms over $F \Gamma $ commutes.
$$\xymatrix@C=3.5cm {
\ar[d]_{T(F\gamma)(i _{A, B})} F( \Pi (A, B))[F\gamma] \ar[r]^{ \theta _{ \Pi (A, B),\gamma}} & F( \Pi (A,
B)[\gamma]) \ar[d]^{{i _{A[\gamma], B[\gamma^+]}}} \\
 \Pi (FA, FB[ \rho ^{-1}_{\Delta, A}])[F\gamma] \ar[r]_{ \Pi ( \theta _{A,  \gamma }^{-1},
\TT( \rho_{\Gamma, A[\gamma]}^{-1})(\theta_{B, \gamma^+}))} &  \Pi (F(A[\gamma]),
F(B[\gamma^+])[\rho_{\Gamma, A[\gamma]}^{-1}])
}$$
Again, it is well-typed by Lemma \ref{lem:theta_lemma}.
\end{lem}\enlargethispage{\baselineskip}
\begin{proof}
The (quite involved) proof appears in Appendix \ref{app:proof_pi}.
\end{proof}

We now resume the inductive proof, and check the inference rules for stability of types under
substitution. We only have to handle the cases for $\I, \Sigma$ and $\Pi$ since $\N_1$ has no
substitution rule.

\begin{description}
\item[$\I$-types] Assume we have
  $ \Delta \vdash a, a' : A$ and $ \Gamma \vdash \gamma : \Delta $.
  Because identity types are extensional, they can be at most one
  isomorphism between identity types, hence
  $\psi_{\I(A, a, a')[\gamma]} = \psi _{\I(A[\gamma], a[\gamma],
    a'[\gamma])}$.
\item[$\Sigma$-types] 
Assume we have
  $ \Delta .A \vdash B$ and $ \Gamma \vdash \gamma : \Delta $. We want
  to prove equality of $ \psi _{ \Sigma (A, B)[\gamma]}$ and
  $ \psi _{ \Sigma (A[\gamma], B[\gamma^+])}$. Since
  $\TT( \varphi _ \Gamma )(s_{A[ \gamma ], B[ \gamma ^+]})$ is an isomorphism, it is equivalent to
  show the equality of
  $\TT( \varphi _ \Gamma )(s_{A[ \gamma ], B[ \gamma ^+]}) \circ \psi _{ \Sigma (A, B)[\gamma]}$
  and
  $\TT( \varphi _ \Gamma )(s_{A[ \gamma ], B[ \gamma ^+]}) \circ \psi _{ \Sigma (A[\gamma],
    B[\gamma^+])}$.

  Calculating yields:
\begin{align*}
&\TT( \varphi _ \Gamma )(s_{A[ \gamma ], B[ \gamma ^+]}) \circ \psi _{ \Sigma (A, B)[ \gamma ]}
\\
 =\ &\TT( \varphi _ \Gamma )(s_{A[ \gamma ], B[ \gamma ^+]}) \circ \TT( \varphi _ \Gamma )( \theta _{
\Sigma (A, B),  \gamma }) \circ \TT(\interp  \gamma )( \psi _{ \Sigma (A, B)}) \\
&\quad\text{(functoriality of $\TT( \varphi _ \Gamma )$)} \\
  =\ & \TT( \varphi _ \Gamma )(s_{A[ \gamma ], B[ \gamma ^+]} \circ  \theta _{ \Sigma (A, B),  \gamma
}) \circ \TT(\interp  \gamma )( \psi _{ \Sigma (A, B)}) \\
&\quad\text{(Lemma \ref{lemma:sigmaTh})} \\
  =\ & \TT( \varphi _ \Gamma )( \Sigma ( \theta _{A,\gamma}, \TT(\rho^{-1}_{\Gamma,A[\gamma]}\circ
\theta _{A,\gamma})( \theta _{B,
\gamma ^+})) \circ \TT(F \gamma )(s_{A, B})) \circ \TT(\interp  \gamma )( \psi _{ \Sigma (A, B)}) \\
&\quad\text{(induction hypothesis on $ \gamma $)} \\
  =\ & \TT( \varphi _ \Gamma )( \Sigma ( \theta _{A,\gamma}, \TT(\rho^{-1}_{\Gamma,A[\gamma]}\circ
\theta _{A,\gamma})( \theta _{B,
\gamma ^+}))) \circ \TT(\interp  \gamma )(\TT( \varphi _ \Delta )(s_{A, B}) \circ  \psi _{ \Sigma
(A, B)}) \\
  =\ & \TT( \varphi _ \Gamma )( \Sigma ( \theta _{A,\gamma}, \TT(\rho_{\Gamma,A[\gamma]}^{-1}\circ
\theta _{A,\gamma})( \theta _{B,
\gamma ^+}))) \circ \TT(\interp  \gamma )( \Sigma ( \psi _{A},  \psi _B)) \\
&\quad\text{(functoriality of $\Sigma(\cdot, \cdot)$ -- Lemmas \ref{lemma:sigmafunctorial} and
\ref{lem:sig_funct_subst})} \\
  =\ & \Sigma (\TT( \varphi _ \Gamma )( \theta _{A,\gamma}) \circ \TT(\interp  \gamma )( \psi _A), \\ & \ \ \ \
\ \ \ \,
\TT(\TT(\interp  \gamma)( \psi _A))(\TT( \varphi _ \Gamma ^+)(\TT(\rho_{\Gamma,A[\gamma]}^{-1}\circ
\theta _{A,\gamma})(
\theta _{B,\gamma^+}))) \circ \TT(\interp  {\gamma^+})( \psi _B)) \\
&\quad\text{(definition of $ \psi _{A[ \gamma ]}$ and functoriality of $\TT$)} \\
=\ & \Sigma \left( \psi _{A[ \gamma ]}, \TT( \rho_{\Gamma, A[\gamma]} ^{-1} \circ  \theta _{A,\gamma} \circ  \varphi _ \Gamma
^+ \circ \TT(\interp  \gamma )( \psi _A))( \theta _{B,\gamma^+}) \circ \TT(\interp{ \gamma ^+})( \psi
_B)\right) \\
&\quad\text{(definition of $ \varphi _{ \Gamma .A [\gamma] }$ + cwf calculations)} \\
=\ & \Sigma \left( \psi _{A[ \gamma ]}, \TT( \varphi _{ \Gamma .A[ \gamma ]})( \theta _{B,\gamma^+}) \circ
\TT(\interp{ \gamma ^+})( \psi _B)\right) \\
=\ & \Sigma ( \psi _{A[ \gamma ]},  \psi _{B[ \gamma^+]}) \\
\end{align*}
\item[$\Pi$-types.] The reasoning is analogous to the case of $\Sigma$ above, using Lemmas
\ref{lemma:pifunctorial}, \ref{lemma:piT} and \ref{lemma:piTh}.
\end{description}

\medskip

\noindent\textbf{Term formation rules.} The term formation rules are those for the introduction of
$0_1$, $\refl(-)$, $\pair, \fst, \snd, \lambda(-)$ and $\ap$.

\begin{description}
\item[Unit] 
We need to prove that $\{\psi_{\bar{1}}\}(0_1) = F0_1 [\varphi_1]$,
where $0_1 \in \Tm_\C(1, \bar{1})$ is defined in the proof of Lemma \ref{lem:unit}. This 
is straightforward by the universal property of the terminal object.
\item[Reflexivity]
Assume that $\Gamma \vdash a = a' : A$. We need to check that 
\[
\{\psi_{\I(A,a, a')}\}(\refl(\interp{a})) = F(\refl(a))[\varphi_\Gamma]
\]
By preservation of $\I$-types we have an iso $$f : F(\I(A,a, a))[\varphi_\Gamma] \cong 
\I(FA[\varphi_\Gamma], Fa[\varphi_\Gamma], Fa[\varphi_\Gamma]),$$ and by the reflection rule we must
have $\{f\}(\{\psi_{\I(A,a, a')}\}(\refl(\interp{a}))) = \{f\}(F\refl(a)[\varphi_\Gamma])$ as they
are both inhabitants of the identity type. 
\item[First projection]
Assume we have $\Gamma \vdash A = A'$, $ \Gamma   \vdash  c = c' :  \Sigma (A, B)$ from which we
  deduce $ \Gamma   \vdash  \fst(A, c) = \fst(A', c') : A$. First, we note that 
$F(\fst(c)) = \fst(\{s_{A, B}\}(Fc))$ by Proposition \ref{prop:preservsigma}. Then, we calculate:

\begin{align*}
  F(\fst(c))[ \varphi _ \Gamma ] &= \fst(\{s_{A, B}\}(F(c)))[ \varphi _ \Gamma ] \\
  &\qquad\text{(definition $\{ \cdot \}$ + interaction $\fst$/substitution)} \\
  &= \fst(\qI[s_{A, B} \circ  \langle \id, F(c) \rangle ][ \varphi _ \Gamma ]) \\
  &\qquad\text{(definition functorial action of $\TT$)} \\
  &= \fst(\{\TT( \varphi _ \Gamma )(s_{A, B})\}(F(c)[ \varphi _ \Gamma ])) \\
  &\qquad\text{(induction hypothesis on $c$)}\\
  &= \fst(\{\TT( \varphi _ \Gamma )(s_{A, B})\}( \{\psi _{ \Sigma (A, B)}\}(\interp c))) \\
  & \qquad \text{(functoriality of $\{ \cdot \}$)}\\
  &= \fst(\{\TT( \varphi _ \Gamma )(s_{A, B}) \circ  \psi _{ \Sigma (A, B)} \}(\interp c))) \\
  &\qquad\text{(Unfolding definition of $ \psi _{ \Sigma (A, B)}$)} \\
  &= \fst(\{ \Sigma ( \psi _A,  \psi _B)\}(\interp c)) \\
  &\qquad\text{(Lemma \ref{lemma:sigmafunctorial})} \\
  &= \qI[ \psi _A \circ  \langle \p, \fst(\qI) \rangle  \circ  \langle \id, \interp c \rangle
\rangle ]  \\
  &= \{ \psi _A\}(\fst(\interp c)) 
\end{align*}
\item[Second projection] Assume we have $\Gamma \vdash A = A', \Gamma.A \vdash B = B',
\Gamma \vdash c = c' : \Sigma(A, B)$ from which we deduce:
\[
\Gamma \vdash \snd(A, B, c) = \snd(A', B', c') : B[\ext{\id_\Gamma, \fst(A, c)}_A]
\]
The calculation follows the same pattern as the one for first projection: we first apply
preservation of the combinators by Proposition \ref{prop:preservsigma}, then calculate.

\begin{align*}
  F(\snd(c))[ \varphi _ \Gamma ] &= \{ \theta _{B,  \langle \id_\Gamma, \fst(c) \rangle_A }\}(\snd(\{s_{A,
B}\}(Fc)))[ \varphi _ \Gamma ]
\\
&\qquad\text{(propagation of $ \varphi _ \Gamma $ and definition of $\TT( \varphi _ \Gamma
)(\cdot)$)} \\
  &= \{\TT( \varphi _ \Gamma )( \theta _{B,  \langle \id_\Gamma, \fst(A,c) \rangle_A }) \}(\snd(\{\TT( \varphi
_ \Gamma )(s_{A, B})\}(Fc[ \varphi _ \Gamma ]))) \\
&\qquad\text{(I.H. on $c$, and definition of $\psi_{\Sigma(A, B)}$)} \\
  &= \{\TT( \varphi _ \Gamma )( \theta _{B,  \langle \id_\Gamma, \fst(A,c) \rangle_A }) \}(\snd(\{ \Sigma (
\psi _A,  \psi _B)\}(\interp c))) \\
&\qquad\text{(unfolding the functorial action of $ \Sigma $)} \\
  &= \{\TT( \varphi _ \Gamma )( \theta _{B,  \langle \id_\Gamma, \fst(A,c) \rangle_A }) \}(\qI[ \psi _B \circ
\langle   \langle \id, \fst(\interp c) \rangle , \snd(\interp c) \rangle ]) \\ 
&\qquad\text{(definition $\TT(\cdot)(\cdot)$)} \\
  &= \{\TT( \varphi _ \Gamma )( \theta _{B,  \langle \id_\Gamma, \fst(A,c) \rangle_A }) \} (\{\TT(\interp{
\langle \id_\Gamma, \fst(A,c) \rangle_A })( \psi _B)\}(\snd (\interp c))) \\
&\qquad\text{(folding definition $ \psi _{ B [ \langle \id_\Gamma, \fst(A, c) \rangle_A ]}$)} \\
  &= \{ \psi _{B[ \langle \id_\Gamma, \fst(A,c) \rangle_A ]}\}(\snd(\interp c)) \\
\end{align*}

\item[Pairing] Assume we have $\Gamma \vdash A = A', \Gamma.A\vdash B = B', \Gamma \vdash a = a' :
A'$, and $\Gamma \vdash b = b' : B[\ext{\id_\Gamma, \fst(A, c)}_A]$. From that, we deduce:
\[
\Gamma \vdash \pair(A, B, a, b) = \pair(A', B', a', b') : \Sigma(A, B)
\]
We start by unfolding the definition of $\psi_{\Sigma(A, B)}$, then calculate:

\begin{align*}
  \{ \psi _{ \Sigma (A, B)}\}(\pair(\interp a, \interp b)) 
  &= \{\TT( \varphi _ \Gamma )(s^{-1}_{A, B})\}\left(\{ \Sigma ( \psi _A,  \psi _B)\}(\pair(\interp
a, \interp b))\right) \\
  &\qquad\text{(Unfolding the definition of $ \Sigma ( \psi _A,  \psi _B)$)}\\
  &= \{\TT( \varphi _ \Gamma )(s^{-1}_{A, B})\}\left(\pair(\{ \psi _A\}(\interp a),  \qI[\psi_B
\circ
\langle  \langle \id, \interp a \rangle , \interp b \rangle ])\right) \\
  &= \{\TT( \varphi _ \Gamma )(s^{-1}_{A, B})\}\left(\pair(\{ \psi _A\}(\interp a), \{\TT( \langle \id,
\interp a \rangle )( \psi _B) \}(\interp b))\right) \\
  &\qquad\text{(definition of $ \psi _{B[ \langle \id, a \rangle ]}$))}\\
  &= \{\TT( \varphi _ \Gamma )(s^{-1}_{A, B})\}\left(\pair(\{ \psi _A\}(\interp a), \{\TT( \varphi _
\Gamma )( \theta^{-1} _{B,  \langle \id_\Gamma, a \rangle_A })\}(\{ \psi _{B[ \langle \id, a \rangle
]}\}(\interp b)))\right)\\
    &\qquad\text{(induction hypothesis on $a$ and $b$)}\\
  &= \{\TT( \varphi _ \Gamma )(s^{-1}_{A, B})\}\left(\pair(Fa[ \varphi _ \Gamma ], \{\TT( \varphi _
\Gamma )( \theta ^{-1}_{B,  \langle \id_\Gamma, a \rangle_A })\}(Fb[ \varphi _ \Gamma ]))\right) \\
  &\qquad\text{(Lemma \ref{lem:coerTT})}\\
  &= \{s^{-1}_{A, B}\}\left(\pair(Fa, \{ \theta ^{-1}_{B,  \langle \id_\Gamma, a \rangle_A
})\}(Fb))\right)[ \varphi _ \Gamma ] \\
  &\qquad\text{(Proposition \ref{prop:preservsigma})}\\
  &= F(\pair(a, b))[ \varphi _ \Gamma ] \\
\end{align*}
\item[Abstraction] Assume we have $\Gamma \vdash A = A'$, $\Gamma.A \vdash b = b' : B$, from which
  we deduce $\Gamma \vdash \lambda(A, b) = \lambda(A', b') : \Pi(A, B)$.

  To limit notational overhead, we omit the first argument of
    lambda abstractions: we often write
    $\lambda(b)$ in place of $\lambda(A, b)$.

We first unfold the definition of $\psi_{\Pi(A, B)}$, and then calculate:

\begin{align*}
  \{ \psi _{ \Pi (A, B)}\}( \lambda (\interp b)) &=
\{\TT( \varphi _ \Gamma )(i_{A, B}^{-1})\}\left(\{ \Pi ( \psi _A^{-1},  \TT( \psi _A^{-1})(\psi
_B))\}( \lambda (\interp b))\right) \\
&\quad\text{(unfolding $\Pi(-,-)$ and long simplifications)}\\
&= \{\TT( \varphi _ \Gamma )(i_{A, B}^{-1})\}\left( \lambda ((\{ \psi _B\}(\interp b))[ \psi
_A^{-1}])\right)\\
&\quad\text{(induction hypothesis on $b$)}\\
&= \{\TT( \varphi _ \Gamma )(i_{A, B}^{-1})\}\left( \lambda ((Fb[ \varphi _{ \Gamma .A}][ \psi
_A^{-1}])\right)\\
&\quad\text{(definition of $ \varphi _{ \Gamma .A}$)}\\
&= \{\TT( \varphi _ \Gamma )(i_{A, B}^{-1})\}\left( \lambda (Fb[ \rho_{\Gamma,A}^{-1}  \circ  \varphi _ \Gamma
^+])\right)\\
&\quad\text{(cwf simplification)}\\
&= \{i_{A, B}^{-1}\}(\lambda(Fb[\rho_{\Gamma, A}^{-1}]))[\varphi_\Gamma]\\
&\quad\text{(Lemma \ref{lemma:pilambda})}\\
&= F( \lambda(b))[ \varphi _ \Gamma ]\\
\end{align*}
\item[Application] Assume that we have $\Gamma \vdash A = A', \Gamma.A \vdash B = B', \Gamma \vdash
c = c' : \Pi(A, B)$, and $\Gamma \vdash a = a' : A$, from which we deduce:
\[
\Gamma \vdash \ap(A, B, c, a) = \ap(A', B', c', a') : B[\ext{\id_\Gamma, a}_A]
\]
As in the previous case, we now drop the $A$ and $B$
annotations in calculations. First we use that $F$ preserves $\Pi$-type (using also
that $\theta_{A, \id} = \id$, which is one of the coherence laws for
pseudo cwf-morphisms), then calculate:

  \begin{align*}
F(\app(c, a))[ \varphi _ \Gamma ]
&= \{ \theta _{B, \langle \id_\Gamma, a \rangle_A} \} \left(\app(\{i_{A, B}\}(Fc), Fa)\right) [ \varphi _ \Gamma
]\\
&\qquad\text{(pushing the substitution by $ \varphi _ \Gamma $ inside)} \\
&= \{\TT( \varphi _ \Gamma )( \theta _{B,  \langle \id_\Gamma, a \rangle_A} )\} \left(\app(\{\TT( \varphi _
\Gamma )(i_{A, B})\}(Fc[ \varphi _ \Gamma ]), Fa[ \varphi _ \Gamma ])\right)\\
&\qquad\text{(induction hypothesis on $c$ and $a$)} \\
&= \{\TT( \varphi _ \Gamma )( \theta _{B,  \langle \id_\Gamma, a \rangle_A} )\} \left(\app(\{\TT( \varphi _
\Gamma )(i_{A, B})\}(\{ \psi _{ \Pi (A, B)}\}(\interp c), \{ \psi _A\}(\interp a))\right) \\
&\qquad\text{(definition of $ \psi _{ \Pi (A, B)}$)} \\
&= \{\TT( \varphi _ \Gamma )( \theta _{B,  \langle \id_\Gamma, a \rangle_A} )\} \left(\app(\{ \Pi ( \psi
_A^{-1}, \TT( \psi _A^{-1})( \psi _B))\}(\interp c), \{ \psi _A\}(\interp a))\right) \\
&\qquad\text{(calculation of $ \Pi ( \psi _A^{-1}, \TT( \psi _A^{-1})( \psi _B))$)} \\
&= \{\TT( \varphi _ \Gamma )( \theta _{B,  \langle \id_\Gamma, a \rangle_A} )\} \left(\qI \left[\TT( \psi
_A^{-1})( \psi _B) \circ \left \langle  \langle \id, \{ \psi _A\}(\interp a) \rangle , \app(\interp
c, \interp a)\right \rangle \right]\right) \\
&\qquad\text{(cwf simplification)} \\
&= \{\TT( \varphi _ \Gamma )( \theta _{B,  \langle \id_\Gamma, a \rangle_A} )\} \left(\qI\left[ \psi _B \circ
\langle  \psi _A^{-1} \circ  \langle \id, \{ \psi _A\}(\interp a) \rangle ,\app(\interp c, \interp
a)\rangle \right]\right)\\
&\qquad\text{(cwf simplification)} \\
&= \{\TT( \varphi _ \Gamma )( \theta _{B,  \langle \id_\Gamma, a \rangle_A} )\} \left(\qI\left[ \psi _B \circ
\langle  \langle \id, \interp a \rangle ,\app(\interp c, \interp a)\rangle \right]\right)\\
&\qquad\text{(folding definitions)} \\
&= \{\TT( \varphi _ \Gamma )( \theta _{B,  \langle \id_\Gamma, a \rangle_A} )\} \left(\{\TT( \langle \id,
\interp a \rangle)(\psi _B)\}(\app(\interp c, \interp a))\right)\\
&\qquad\text{(definition of $\psi_{B[\ext{\id_\Gamma, a}_A}$)}\\
&= \{ \psi _{B[ \langle \id_\Gamma, a \rangle_A ]}\}(\app(\interp c, \interp a)))\\
  \end{align*}
\end{description}

\medskip

\noindent\textbf{Conversion, and substitution on terms.} The last rules left to check are the
conversion rules, and the substitution on terms. We do not detail them, as they are all immediate
consequences of the corresponding rules for equality and the substitution on terms in the cwf
structure.

\subsubsection{Uniqueness of $\varphi$ and $\psi$}
We resume the uniqueness proof from Section \ref{sec:cwf_uniq}.

\medskip

\noindent\textbf{Unit type.} Since $1.\N_1  \cong  1$, uniqueness follows from the terminal object universal
property.

\medskip

\noindent\textbf{Identity types.} We need to show
$ \psi' _{\I(A, a, a')} =  \psi _{\I(A, a, a')} :  \Gamma .\I(A, a, a')   \rightarrow    \Gamma .F(\I(A, a,
a'))[ \varphi _ \Gamma ]$.
By post-composing with the coherence isomorphism
$F(\I(A, a, a'))  \cong _{F \Gamma } \I(FA, Fa, Fa')$, we get a morphism
between identity types. In an extensional type theory, identity types
are either empty or singletons, thus there is at most one morphism
between two identity types (which is an isomorphism). This implies that
$ \psi _{\I(A, a, a')} =  \psi '_{\I(A, a, a')}$.

\medskip

\noindent\textbf{$ \Sigma $-types.} By induction, we assume that
  $ \varphi _{ \Gamma .A.B} = \varphi '_{ \Gamma .A.B}$.  By
  naturality of $ \varphi '$, we have
  $ \varphi '_{ \Sigma (A, B)} = F( \chi _{A, B}^{-1})\circ \varphi
  '_{ \Gamma .A.B} \circ \chi _{A, B} = \varphi _{ \Gamma . \Sigma (A,
    B)}$.
  Because $ \varphi $ is also natural, we can derive a similar equation, hence
  $ \psi _{ \Sigma (A, B)} = \psi '_{ \Sigma (A, B)}$.

\medskip

\noindent\textbf{$ \Pi $-types.} By induction we
  assume $ \varphi _{ \Gamma .A.B} = \varphi '_{ \Gamma .A.B}$. It also follows from induction
hypothesis that $\varphi_\Gamma = \varphi'_\Gamma$, $\psi_A = \psi'_A$ and $\psi_B = \psi'_B$.

%
%

Let $\ev_{A, B}$ be the evaluation map, a morphism in $\Gamma.A$:
\[
\ev_{A, B} = \ext{\p, \app(\qI,\qI[\p])} : \Pi(A, B)[\p] \to B
\]
Proposition \ref{prop:chos_dep_prod} entails:
\begin{lem}\label{lem:uniproppi}
Take $B \in \Ty_\C(\Gamma.A)$ in any cwf $\C$ with $\Pi$-types. The only automorphism $ f $ of $ \Pi (A, B)$ (in
  $\TT  \Gamma $) such that $T(\p)( f ) :  \Gamma .A. \Pi (A, B)[\p]  \cong   \Gamma .A. \Pi (A, B)[\p]$ satisfies
$\ev_{A,B} \circ T(\p)( f ) = \ev_{A,B}$ is the identity.
\end{lem}

We will exploit this, and show that $\psi _{ \Pi (A, B)}^{-1} \circ \psi '_{ \Pi (A, B)}  $ satisfies the
condition. But first, we prove that the $\psi$ component of a pseudo cwf-transformation from
$\interp{-}$ to $F$ automatically preserves evaluation, in the following sense.

\begin{lem}\label{lem:auxpsi}
Let $(\varphi, \psi)$ be any pseudo cwf-transformation from $\interp{-}$ to $F$. Then, we have:
\[
\ev'_{A,B} \circ \TT(\p)(\psi_{\Pi(A, B)}) = \ev_{A, B} : \interp{\Gamma.A.\Pi(A,B)[\p]} \to
\interp{\Gamma.A.B}
\]
where we use an alternative evaluation map:
\begin{eqnarray*}
\ev'_{A,B} &=& \varphi_{\Gamma.A.B}^{-1} \circ F(\ev_{A, B}) \circ \rho_{\Gamma.A, \Pi(A,B)[\p]}^{-1}
\circ \theta_{\Pi(A, B), \p} \circ \varphi_{\Gamma.A}^+\\
&:& \interp{\Gamma.A}.F(\Pi(A,B))[\varphi_\Gamma \circ \p] \to \interp{\Gamma.A.B}
\end{eqnarray*}
\end{lem}
\begin{proof}
We calculate:
\begin{align*}
  &F(\ev_{A,B}) \circ  \rho_{\Gamma.A, \Pi(A,B)[\p]} ^{-1} \circ  \theta _{ \Pi (A, B), \p} \circ
\varphi _{ \Gamma .A}^+ \circ T(\p)( \psi_{ \Pi (A, B)}) \\
  =\ &F(\ev_{A,B}) \circ  \rho_{\Gamma.A, \Pi(A,B)[\p]} ^{-1} \circ  \varphi _{ \Gamma .A}^+ \circ
T( \varphi _{ \Gamma .A})( \theta_{\Pi (A, B), \p} ) \circ T(\p)( \psi_{ \Pi (A, B)}) \\
&\qquad\text{(Lemma \ref{lem:transcoh})} \\
  =\ &F(\ev) \circ  \rho ^{-1}_{\Gamma.A, \Pi(A,B)[\p]} \circ  \varphi _{ \Gamma .A}^+ \circ  \psi _{ \Pi (A, B)[\p]} \\
&\qquad\text{(Coherence of pseudo cwf-transformations)}\\
  = \ &F(\ev) \circ  \varphi _{ \Gamma .A. \Pi (A, B)[\p]} \\
&\qquad\text{(Naturality of $\varphi$)}\\
  =\ & \varphi_{ \Gamma .A.B} \circ \ev_{A,B}
\end{align*}
Importantly, this is proved not with the inductive definition of $(\varphi, \psi)$, but only using
general properties of pseudo cwf-transformations.
\end{proof}

Using that both $\psi_{\Pi(A, B)}$ and $\psi'_{\Pi(A, B)}$ satisfy the property above, their
equality follows easily. We calculate:

\begin{align*}
&\ \ev_{A, B} \circ \TT(\p)(\psi_{\Pi(A, B)}^{-1}) \circ \TT(\p)(\psi'_{\Pi(A,B)}) \\
&\qquad\text{(Lemma \ref{lem:auxpsi} on $\psi_{\Pi(A, B)}$)}\\
=&\ \ev'_{A, B} \circ \TT(\p)(\psi'_{\Pi(A,B)})\\
&\qquad\text{(Lemma \ref{lem:auxpsi} on $\psi_{\Pi(A, B)}'$)}\\
=&\ \ev_{A, B}
\end{align*}

Hence, $\psi_{\Pi(A, B)} = \psi'_{\Pi(A, B)}$ by Lemma \ref{lem:uniproppi}.

%
\subsection{The free lccc }

Let $\textbf{LCC}$ be the 2-category of lcccs. Since biequivalences
preserve bi-initiality, the biequivalence of \cite{clairambault:mscs}
$\Cwf^{ \Sigma , \Pi , I}_d \simeq \textbf{LCC}$ allows us to turn the
bi-initial cwf into a bi-initial LCCC:
\begin{thm}\label{bifree-lccc}
The category of contexts of $\T^{\I,\N_1, \Sigma , \Pi}$ is a bifree lccc on one object, that is, it is bi-initial in $\textbf{LCC}^o$.
\end{thm}

\section{Conclusion}

We have shown that a version of Martin-Löf Type Theory gives rise to the \emph{free} cwf,
with and without $\I, \N_1, \Sigma$ and $\Pi$. We have proved this freeness result both in a
1-categorical sense (with respect to \emph{strict cwf-morphisms}), and in a $2$-categorical sense
(with respect to \emph{pseudo cwf-morphisms}). It follows that the category of
contexts of our type theory $\T^{\I,\N_1, \Sigma , \Pi}$ is a bifree lccc. We also proved that
equality is undecidable in $\T^{\I,\Pi}$ (improving slightly on the folklore result),
hence showing undecidability of equality in a bifree lccc.
There is only one bifree lccc up to equivalence, so in that sense $\T^{\I,\N_1, \Sigma , \Pi}$ is
\emph{the} bifree lccc. 
However, note that the undecidability statement is only about our
particular presentation of the bifree lccc, and not about an arbitrary
one.
We could introduce a notion of recursively presented lccc and ask the more general question whether
an arbitrary such recursively presented bifree lccc has undecidable equality, but we will leave that for future work.

\bibliographystyle{plain}
\bibliography{refs}

\appendix

\section{Combinators for pseudo cwf-morphisms and results from \cite{clairambault:mscs}}
\label{app:pseudo_mor}

\subsection{Pseudo cwf-morphisms}
We first give the full definition of pseudo cwf-morphisms along with the coherence and
naturality laws that were only sketched in the main text, and we recall a few important results
concerning their manipulations.

First, we recall a notation from \cite{clairambault:mscs}. Let $\C$ be a cwf. In Section
\ref{sec:cwfs}, we introduced a functor:
\[
\indexed{T} : \C^{\op} \to \Cat
\]
which in particular associates, to any object $\Gamma$ of $\C$, a category $\indexed{T}(\Gamma)$. We
introduced the notation $f : A \iso_\Gamma B$ to mean that $f$ is an invertible map from
$A$ to $B$ in $\indexed{T}{\Gamma}$. Then, for $a : \Gamma \vdash A$, we also introduced
$\{f\}(a) : \Gamma \vdash B$ for the \emph{coercion} of $a$ to type $B$.
In that case, whenever $b = \{f\}(a)$, we introduce the new notation
\[
b =_f a
\]
meaning that $a$ and $b$ are the same up to coercion.

We are now in position to give the full definition of a pseudo cwf-morphism.

\begin{defi}
A \textbf{pseudo cwf-morphism} from $(\C, T)$ to $(\C', T')$ is a pair $(F,
\sigma)$ where:
\begin{itemize}
\item $F: \C \to \C'$ is a functor,
\item For each context $\Gamma$ in $\C$, $\sigma_\Gamma$ is a $\Fam$-morphism from $T \Gamma$ to $T'
(F \Gamma)$. We will write $\sigma_\Gamma(A) \in \Ty'(F\Gamma)$, where $A \in \Ty(\Gamma)$, for the type
component and $\sigma^A_\Gamma(a) : F\Gamma \vdash ' \sigma_\Gamma(A)$, where $a : \Gamma \vdash A$,
for the term component of this morphism.
\end{itemize}
The following preservation properties must be satisfied:
\begin{itemize}
\item Substitution is preserved: For each substitution $\gamma : \Delta \to \Gamma$ in $\C$ and $A\in
\Ty(\Gamma)$, there is an isomorphism of types
$\theta_{A, \gamma} : \sigma_\Gamma(A)[F\gamma] \to \sigma_\Delta(A[\gamma])$ such that substitution
in terms is also preserved, that is,
$\sigma_\Delta^{A[\gamma]}(a[\gamma]) =_{\theta_{A,\gamma}} \sigma_\Gamma^A(a) [F \gamma]$.
\item The terminal object is preserved: $F\nilc$ is terminal, let $!_F : \nilc \to F\nilc$ be the
isomorphism.
\item Context comprehension is preserved: The context $F(\Gamma\cext A)$, along with the projections
$F(\p_{\Gamma,A})$ and $\{\theta_{A, \p}^{-1}\}(\sigma_{\Gamma\cext A}^{A[\p]}(\qI_{\Gamma,A}))$,
is a context comprehension of $F\Gamma$ and $\sigma_{\Gamma}(A)$. Note that the universal property
of context comprehension provides a unique
isomorphism $\rho_{\Gamma, A}: F(\Gamma\cext A) \to F\Gamma\cext \sigma_\Gamma(A)$ which preserves
projections in the following sense:
\begin{align}
F(\p_A) &= \p_{\sigma_\Gamma(A)} \rho_{\Gamma, A}\label{eq:presp}\tag{a}\\
\sigma_{\Gamma\cext A}^{A[\p]}(\qI_A) &=_{\theta_{A, \p}} \qI_{\sigma_\Gamma A}[\rho_{\Gamma,
A}]\label{eq:presq}\tag{b}
\end{align}
\end{itemize}
These data must satisfy naturality and coherence laws which amount to the fact that if we extend
$\sigma_\Gamma$ to a functor
$\indexed{\sigma}_\Gamma : \indexed{T}(\Gamma) \to \indexed{T'}F(\Gamma)$, then $\indexed{\sigma}$
is a pseudonatural transformation from $\indexed{T}$ to $\indexed{T'}F$.  This functor is defined by
$\indexed{\sigma}_\Gamma(A) = \sigma_\Gamma(A)$ on an object $A$ and $\indexed{\sigma}_\Gamma(f) =
\rho_{\Gamma, B} F(f) \rho^{-1}_{\Gamma, A}$ on a morphism $f: A\to B$.

More explicitely, pseudonaturality of $\indexed{\sigma}$ amounts to the following coherence and
naturality laws.
\begin{itemize}
\item \emph{Identity.} For all $A\in \Ty(\Gamma)$, we have $\theta_{A,\id} = \id_{F\Gamma \cext
\sigma_\Gamma(A)}$,
\item \emph{Coherence.} For all $\delta: \Xi \to \Delta$ and $\gamma : \Delta \to \Gamma$, the
following diagram commutes.
\[
\xymatrix
{
F \Xi \cext \sigma_\Gamma(A)[F(\gamma \delta)]
\ar[dr]_{\indexed{T'}(F\delta)(\theta_{A,\gamma})}
\ar[rr]^{\theta_{A,\gamma \delta}}
&&
F \Xi \cext \sigma_\Xi(A[\gamma\delta])\\
&F \Xi \cext \sigma_\Delta(A[\gamma])[F(\delta)]
\ar[ur]_{\theta_{A[\gamma],\delta}}
}
\]
\item \emph{Naturality.} For all $\delta: \Delta \to \Gamma$ in $\C$, $A, B\in \Ty(\Gamma)$ and $f:
A \to B$ in $\indexed{T}(\Gamma)$, the following
diagram commutes in $\indexed{T'}(F\Delta)$.
\[
\xymatrix@C=60pt{
\sigma_\Gamma(A)[F\delta]
        \ar[r]^{\theta_{A, \delta}}
        \ar[d]^{\indexed{T'}(F\delta)(\indexed{\sigma}_\Gamma(f))}&
\sigma_\Delta(A[\delta])
        \ar[d]^{\indexed{\sigma}_\Delta(\indexed{T}(\delta)(f))}\\
\sigma_\Gamma(B)[F\delta]
        \ar[r]^{\theta_{B, \delta}}&
\sigma_\Delta(B[\delta])
}
\]
\end{itemize}
\end{defi}

This concludes the full definition of pseudo cwf-morphisms. We now recall 
a few key properties of those that are used in the course of the paper. For
the proofs we refer to \cite{clairambault:mscs}.

Firstly, it follows from the definition above that substitution extension is preserved up to
isomorphism.

\begin{prop}
All pseudo cwf-morphisms $(F, \sigma)$ from $(\C, T)$ to $(\C', T')$ preserve substitution extension
in the sense that, if $\delta : \Delta \to \Gamma$ in $\C$ and $a:\Delta \vdash A[\delta]$, then
\[
F(\subst{\delta,a}) =
\rho_{\Gamma,A}^{-1}\subst{F\delta,\applyopen{\theta_{A,\delta}^{-1}}{(\sigma_\Delta^{A[\delta]}(a))}}
\]
\label{prop:pres_subst_ext}
\end{prop}
\begin{proof}
Proposition 3.2 of \cite{clairambault:mscs}.
\end{proof}

It is convenient to also have a specialized version of the proposition above, in the case where the
substitution has the form $\subst{\gamma \p, \qI} : \Delta \cext A[\gamma] \to \Gamma \cext A$,
\emph{i.e.} it is the \emph{lifting} of a substitution $\gamma : \Delta \to \Gamma$ in the presence
of an additional type $A \in \Ty(\Gamma)$. We get:

\begin{lem}
Let $(F, \sigma): (\C, T) \to (\C', T')$ be a pseudo cwf-morphism with families of isomorphisms
$\theta$ and $\rho$. Then for any $\delta : \Delta \to \Gamma$ in $\C$ and type $A\in \Ty(\Gamma)$,
we have:
\[
F(\subst{\delta \p, \qI}) = \rho_{\Gamma, A}^{-1} \subst{F(\delta)\p, \qI} \theta_{A, \delta}^{-1}
\rho_{\Delta, A[\delta]}
\]
\label{lem:pres_lift_sub}
\end{lem}
\begin{proof}
Lemma A.2 of \cite{clairambault:mscs}.
\end{proof}

Finally, we mention a technical lemma stating a compatibility between substitutions and coercions.

\begin{lem}
Let $(\C, T)$ be a cwf. Let $\delta : \Delta \to \Gamma$ be a substitution, $f : A \iso_\Gamma A'$
an isomorphism of types in $\Ty(\Gamma)$ and
$a:\Gamma\vdash A$ be a term. Then:
\[
(\{f\}(a))[\delta] = \{\indexed{T}(\delta)(f)\}(a[\delta])
\]
\label{lem:comp_subst_coer}
\end{lem}
\begin{proof}
Lemma A.1 of \cite{clairambault:mscs}.
\end{proof}

\subsection{Preservation of type formers}

Now, we recall some material from \cite{clairambault:mscs} about the preservation of $\Pi$-types by
pseudo cwf-morphisms. In particular, we recall a more abstract characterization of preservation of
$\Pi$-types, based on a universal property satisfied by $\Pi$-types: that of a
\emph{dependent product diagram}.

If $g:A\to B$ and $f:B \to C$ are morphisms in a category $\C$, a dependent product of $g$ along $f$
is a
diagram of the form:
\[
\xymatrix{
&P      \ar@/_/[dl]_{ev}
        \ar[d]
        \ar[r]
        \pb{315}&
D       \ar[d]^{\Pi_f(g)}\\
A       \ar[r]^g&
B       \ar[r]^f&
C
}
\]
which is universal among all such diagrams in $g$ and $f$ in the following sense:
\[
\xymatrix{
&P'     \ar@/_/[ddl]
        \ar@/_/[dd]
        \ar@{.>}[d]
        \ar[r]
        \pb{315}&
D'      \ar@/_/[dd]
        \ar@{.>}[d]\\
&P      \ar@/_/[dl]
        \ar[d]
        \ar[r]
        \pb{315}&
D       \ar[d]\\
A       \ar[r]&
B       \ar[r]&
C
}
\]
It follows from the universal property that dependent products of $g$ along $f$ are unique up to
isomorphism. First, we observe that $\Pi$-types indeed yield dependent product diagrams.

\begin{prop}\label{prop:chos_dep_prod}
Let $(\C, T)$ be a cwf supporting $\Pi$-types, let $\Gamma$ be a context in $\C$, let $A\in
\Ty(\Gamma)$ and $B\in \Ty(\Gamma\cext A)$, then
the following diagram is a dependent product diagram, where $ev_{A, B} = \subst{\p_{\Pi(A,
B)[\p_A]}, \ap(\qI_{\Pi(A, B)[\p_A]}, \qI_A[\p_{\Pi(A, B)[\p_A]}])}$.
\[
\xymatrix{
&\Gamma\cext A\cext \Pi(A, B)[\p_A]
        \ar@/_/[dl]_{ev_{A, B}}
        \ar[d]^{\p_{\Pi(A, B)[\p_A]}}
        \ar[rr]^{\subst{\p_A\p_{\Pi(A, B)[\p_A]}, \qI_{\Pi(A, B)[\p_A]}}}
        \pb{315}&~~~~~~~~~~~~~~~~&
\Gamma\cext \Pi(A, B)
        \ar[d]^{\p_{\Pi(A, B)}}\\
\Gamma\cext A \cext B
        \ar[r]^{\p_B}&
\Gamma\cext A
        \ar[rr]^{\p_A}&&
\Gamma
}
\]
It is referred to as the \emph{chosen} dependent product of $\p_B$ along $\p_A$.
\end{prop}
\begin{proof}
Proposition 4.6 in \cite{clairambault:mscs}.
\end{proof}

With this in place, we can recall from \cite{clairambault:mscs} the characterization of preservation
of $\Pi$-types in terms of preservation of dependent product diagrams. 

\begin{lem}\label{lem:carac_pres_pi}
Let $(\C, T)$ and $(\C', T')$ be cwfs supporting $\Pi$-types, and $(F, \sigma)$ be a pseudo
cwf-morphism from $(\C, T)$ to $(\C', T')$. Then
$(F, \sigma)$ preserves $\Pi$-types if and only if the image of any chosen dependent product diagram
is a dependent product diagram.
\label{lem:pres_chosen}
\end{lem}
\begin{proof}
Lemma 4.7 in \cite{clairambault:mscs}. 
\end{proof}\nobreak\enlargethispage{\baselineskip}
\subsection{Preservation of isomorphisms by type formers}

We recall only one lemma from \cite{clairambault:mscs} which is not covered by the development in
the main text: that identity types preserve isomorphisms.

\begin{lem}\label{lem:prop_id_iso}
For any $A, A'\in
\Ty_\C(\Gamma)$, $B, B'\in \Ty_\C(\Gamma\cext A)$, 
if $f : A\iso_\Gamma A'$ and $a, a' \in \Gamma\vdash A$, then $\Id(A,a, a') \iso_\Gamma
\Id(A',\applyopen{f}{(a)}, \applyopen{f}{(a')})$
\end{lem}
\begin{proof}
It is one case of the Lemma B.1 from \cite{clairambault:mscs}.
\end{proof}

\section{On pseudo cwf-transformations (erratum for \cite{clairambault:mscs})}
\label{app:pseudocwftrans}

In \cite{clairambault:mscs}, pseudo cwf-transformations ($2$-cells in the $2$-category of cwfs) are
defined as follows.

\begin{defi}\label{def:pseudo2}[Pseudo cwf-transformation -- version of \cite{clairambault:mscs}]
Let $F$ and $G$ be cwf-morphisms from $(\C, T)$ to $(\C', T')$. A \emph{pseudo
cwf-transformation} from  $F$ to
$G$ is a pair $(\varphi, \psi)$ where $\varphi: F \Rightarrow G$ is a natural transformation, and $\psi_{A} : FA \to GA[\varphi_\Delta]$ is a morphism in $\mathbf{T'}(F\Delta)$
for each $\Delta$ in $\C$ and $A\in \Ty_\C(\Delta)$. Furthermore, $\psi_A$ is natural in $A$
and the following diagram commutes:
\[
\xymatrix@C=60pt@R=20pt{
FA[F\gamma]
\ar[r]^{\mathbf{T'}(F\gamma)(\psi_{A})}
\ar[d]^{\theta^F_{A, \gamma}}&
GA[\varphi_\Delta F(\gamma)]
\ar[d]^{\mathbf{T'}(\varphi_\Gamma)(\theta^G_{A, \gamma})}\\
F(A[\gamma])
\ar[r]_{\psi_{A[\gamma]}}&
G(A[\gamma])[\varphi_\Gamma]
}                                                                                                                                                     
\]
Here $\theta$ and $\theta'$ are the isomorphisms witnessing preservation of substitution in types
in the definition of pseudo cwf-morphisms.
\end{defi}

When working on the present paper, we discovered a shortcoming of this definition:
the component
$\psi$ is not constrained enough by $\varphi$. This causes a mismatch with the $2$-cells in
$\mathbf{LCC}$ (where only the $\varphi$ remains), and
as a consequence the family of cwf-transformations $\epsilon$ used in the biequivalence (see
\cite{clairambault:mscs}) fails to satisfy the required condition of
pseudonatural transformations.

What is missing from the definition of pseudo cwf-transformation is the following coherence diagram which must commute for the biequivalence to hold:
\[
\xymatrix{
F(\Delta.A) \ar[rr]^{\varphi_{\Delta.A}}\ar[d]_{\rho^F_{\Delta, A}}&&
G(\Delta.A)\ar[d]^{\rho^G_{\Delta, A}}\\
F\Delta.FA\ar[r]^{\psi_{A}}&F\Delta.FA[\varphi_\Delta]\ar[r]^{\varphi_\Delta^+}& G\Delta.GA
}
\]
This diagram shows that $\psi$ can be defined from $\varphi$. Hence we could simply define pseudo cwf-transformations as natural transformations
$\varphi : F \Rightarrow G$. However, we have not done so, because pseudo
cwf-transformations are most naturally presented
with the $\psi$, reflecting the second component of cwfs and cwf-morphisms. Moreover, in our proof
of bifreeness, the construction of
the unique cwf-transformation between the interpretation and an arbitrary pseudo cwf-functor
naturally constructs $\varphi$ and $\psi$ by mutual induction.

With the addition of the coherence diagram above, the naturality requirement and the 
coherence diagram of Definition \ref{def:pseudo2} become
redundant, as we establish here. The following lemma is a mild generalization of Lemma 5.6 of
\cite{clairambault:mscs}.

\begin{lem}\label{lem:transcoh}
Let $F, G : \C \to \C'$ be pseudo cwf-morphisms, and let $(\varphi,\psi)$ be a
pseudo cwf-transformation from $F$ to $G$, in the sense of Definition \ref{def:pseudo1}. Then,
$(\varphi, \psi)$ is also a pseudo cwf-transformation in the sense of Definition \ref{def:pseudo2},
\emph{i.e.} it satisfies the coherence law:
\[
\xymatrix@C=60pt@R=20pt{
FA[F\gamma]
\ar[r]^{\mathbf{T'}(F\gamma)(\psi_{A})}
\ar[d]^{\theta_{A, \gamma}}&
GA[\varphi_\Delta F(\gamma)]
\ar[d]^{\mathbf{T'}(\varphi_\Gamma)(\theta'_{A, \gamma})}\\
F(A[\gamma])
\ar[r]_{\psi_{A[\gamma]}}&
G(A[\gamma])[\varphi_\Gamma]
}                                                                            
\]
\end{lem}
\begin{proof}
We first check that $\psi_A$ is natural in $A$. More explicitely, recall from 
\cite{clairambault:mscs} that each cwf-transformation $F : \C
\to \C'$ induces, for any context $\Gamma$ of $\C$, a functor:
\[
F_\Gamma : \TT(\Gamma) \to \TT'(F\Gamma)
\]
Its action on types is obvious. Recall that a morphism $f: A \to B$ in $\TT(\Gamma)$ is a
morphism $f : \Gamma.A \to \Gamma.B$ in $\C$ such that $\p \circ f = \p$. Its action on $F_\Gamma$
is simply:
\[
\rho^F_{\Gamma, B} \circ F(f) \circ (\rho^F_{\Gamma, A})^{-1} : F\Gamma . FA \to F\Gamma . FB
\]
The naturality of $\psi_A$ in $A$ follows directly from the naturality of $\varphi$.

\medskip

We also need to check that the coherence law of Definition \ref{def:pseudo2} holds.
We follow the proof of Lemma 5.6 in \cite{clairambault:mscs}, and consider the following composition
of squares:
\[
\xymatrix@C=40pt{
F\Gamma. G(A[\gamma])[\varphi_\Gamma]
        \pb{315}
       \ar[r]^{\varphi_\Gamma^+}
       \ar[d]^{\p}&
G\Gamma. G(A[\gamma])
        \pb{315}
       \ar[rr]^{\rho_{\Delta, A}^G \circ G(\gamma^+) \circ (\rho_{\Gamma, A[\gamma]}^G)^{-1}}
       \ar[d]^\p&&
G\Delta. G(A)
       \ar[d]^\p\\
F\Gamma        \ar[r]_{\varphi_\Gamma}&
G\Gamma        \ar[rr]_{G\gamma}&&
G\Delta
}
\]
The left hand side square is a pullback -- the standard substitution pullback of $\p_{G(A[\gamma])}$ along
$\varphi_\Gamma$. In \cite{clairambault:mscs}, it is noted that the right hand side square is also a
pullback, as the image of a substitution pullback through $G$; which is there assumed to preserve
pullbacks. For us though $F$ does not in general preserve pullbacks, but it preserves this one.
Indeed, by the commutation property of Lemma \ref{lem:theta_lemma} it is straightforward to prove it to be isomorphic to the
substitution pullback of $\p_{GA}$ along $G\gamma$.

Therefore, the composition of the two squares is a pullback as well.
Once we have established this, the proof follows exactly as in the proof
of Lemma 5.6 in \cite{clairambault:mscs}. We exploit that the
two paths $\TT(\varphi_\Gamma)(\theta^G_{A, \gamma}) \circ \TT(F\gamma)(\psi_{
A})$ and $\psi_{A[\gamma]}\circ \theta^F_{A, \gamma}$ of the coherence
diagram behave in the same way with respect to this pullback, and therefore are equal by the
universal property. The calculations are given in detail in \cite{clairambault:mscs}, so we do not repeat
them here.
\end{proof}

Thus all pseudo cwf-transformations in the sense of Definition \ref{def:pseudo1} are
also pseudo cwf-transformations in the sense of Definition \ref{def:pseudo2}.
Moreover, all pseudo cwf-transformations used in the biequivalence \cite{clairambault:mscs}
trivially obey this stronger condition. In fact, all the pseudo cwf-transformations $(\varphi,
\psi)$ used in the biequivalence were defined by their $\varphi$ component, whereas the $\psi$
component was defined \emph{a posteriori} via the equation of Definition \ref{def:pseudo1}.

\section{Proof of Lemma \ref{lemma:piTh}}
\label{app:proof_pi}

The proof of Lemma \ref{lemma:piTh} uses the notion of dependent product diagram and in
particular the corresponding universal property -- both paths around the diagram will be proved to
preserve the structure of some dependent product diagrams. Their equality will immediately follow
from the uniqueness component of the universal property. We now inspect in turn all four morphisms
of the diagram of Lemma \ref{lemma:piTh}, and prove that they preserve dependent product
structure.

In the remainder of this section, we consider cwfs $\C, \C'$ supporting $\Pi$-types, $F$ from $\C$
to $\C'$ preserving $\Pi$-types, a substitution $\gamma : \Gamma \to \Delta$ of $\C$ and types $A\in
\Ty_\C(\Delta), B\in \Ty_\C(\Delta.A)$. We first note:

\begin{lem}\label{lem:dia0}
The following is a dependent product diagram.
\[
\xymatrix{
&F\Delta.FA.F(\Pi(A,B))[\p]      \ar@/_/[dl]_{\ev_{F(\Pi(A, B))}}
        \ar[d]^\p
        \ar[r]^{\p^+}
        \pb{315}&
F\Delta.F\Pi(A,B)       \ar[d]^{\p}\\
F\Delta.FA.FB[\rho_{\Delta, A}^{-1}]       \ar[r]^\p&
F\Delta.FA       \ar[r]^\p&
F\Delta
}
\]
where $\ev_{F(\Pi(A,B))} = \rho_{\Delta,A}^+ \circ \rho_{\Delta.A, B} \circ F(\ev_{A, B}) \circ
\rho_{\Delta.A, \Pi(A, B)[\p]}^{-1} \circ \theta_{\Pi(A, B), \p} \circ (\rho_{\Delta, A}^{-1})^+$.

This means that there is a unique isomorphism to the chosen dependent product diagram of $FA$ and
$FB[\rho_{\Delta, A}^{-1}]$, which is given by the morphism:
\[
i_{A, B} : F(\Pi(A, B)) \to \Pi(FA, FB[\rho_{\Delta, A}^{-1}])
\]
involved in the definition of pseudo cwf-morphisms preserving $\Pi$-types. The fact that it yields a
morphism of dependent product diagrams means that we also have:
\[
\ev_{FA, FB[\rho_{\Delta, A}^{-1}]} \circ \TT(\p)(i_{A, B}) = \ev_{F(\Pi(A, B))}
\]
\end{lem}
\begin{proof}
By Lemma \ref{lem:carac_pres_pi}, the image by $F$ of the chosen dependent product diagram of
$A$ and $B$ is a dependent product diagram. From this diagram we can obtain the diagram above by applying
structural isomorphisms $\rho$ and $\theta$ on the nodes. Being obtained by transporting a
dependent product diagram along isomorphisms, it is itself a dependent product diagram.
Its evaluation morphism, $\ev_{F(\Pi(A, B))}$, is obtained by going through the isomorphisms.
The fact that $i_{A, B}$
corresponds to the canonical dependent product diagram morphism is a direct verification, using 
the universal property of dependent product diagrams.
\end{proof}

From that follows immediately:

\begin{lem}\label{lem:dia1}
The following is a morphism between two dependent product diagrams:
\[
\scalebox{0.8}{
\xymatrix@R=30pt{
&F\Gamma.FA[F\gamma].F(\Pi(A,B))[(F\gamma)\circ\p]      \ar@/_/[dl]_{\TT(F\gamma)(\ev_{F(\Pi(A,
B))})}
	\ar@/^2pc/[ddd]^{\TT((F\gamma)\circ \p)(i_{A, B})}
        \ar[d]^\p
        \ar[r]^{\p^+}
        \pb{315}&
F\Gamma.F\Pi(A,B)[F\gamma]       
	\ar[d]^{\p}
	\ar@/^2pc/[ddd]^{\TT(F\gamma)(i_{A, B})}\\
F\Gamma.FA[F\gamma].FB[\rho_{\Delta, A}^{-1}][(F\gamma)^+]       \ar[r]^\p&
F\Gamma.FA[F\gamma]       \ar[r]^\p&
F\Gamma\\\\
&F\Gamma.FA[F\gamma].\Pi(FA,FB[\rho_{\Delta,A}^{-1}])[(F\gamma)\circ\p]      \ar@/_/[dl]_{\ev_{FA[F\gamma],
FB[(F\gamma)^+]}}
        \ar[d]^\p
        \ar[r]^{\p^+}
        \pb{315}&
F\Gamma.\Pi(FA,FB[\rho_{\Delta, A}^{-1}])[F\gamma]       \ar[d]^{\p}\\
F\Gamma.FA[F\gamma].FB[\rho_{\Delta, A}^{-1}][(F\gamma)^+]       \ar[r]^\p&
F\Gamma.FA[F\gamma]       \ar[r]^\p&
F\Gamma
}}
\]
where all arrows not explicitely displayed are identities.
\end{lem}
\begin{proof}
This diagram is obtained by pulling back that of Lemma \ref{lem:dia0} along $F\gamma$ -- it is
straightforward that this operation preserves dependent product diagrams.
\end{proof}

Thus we have proved that the left hand side and the right hand side (instantiating Lemma
\ref{lem:dia0} with $A[\gamma], B[\gamma^+]$) maps of Lemma \ref{lemma:piTh}
correspond as required to morphisms of dependent product diagrams. This remains to be done for the
upper and lower maps. We start with the lower map.

\begin{lem}\label{lem:dia2}
The following is a morphism between two dependent product diagrams:
\[
\scalebox{0.7}{
\xymatrix@R=30pt{
&F\Gamma.FA[F\gamma].\Pi(FA,FB[\rho_{\Delta,A}^{-1}])[(F\gamma)\circ\p]
\ar@/_/[dl]_{\ev_{FA[F\gamma],
FB[(F\gamma)^+]}}
	\ar@/^2pc/[ddd]_{\TT(\p)(\Pi(\theta_{A, \gamma}^{-1}, \TT(\rho_{\Gamma,
A[\gamma]}^{-1})(\theta_{B, \gamma^+})))\circ \theta_{A, \gamma}^+}
        \ar[d]^\p
        \ar[r]^{\p^+}
        \pb{315}&
F\Gamma.\Pi(FA,FB[\rho_{\Delta, A}^{-1}])[F\gamma]       
	\ar[d]^{\p}
	\ar@/^2pc/[ddd]_{\Pi(\theta_{A, \gamma}^{-1}, \TT(\rho_{\Gamma,
A[\gamma]}^{-1})(\theta_{B, \gamma^+}))}\\
F\Gamma.FA[F\gamma].FB[\rho_{\Delta, A}^{-1}][(F\gamma)^+]      
	\ar[r]^\p
	\ar@/^2pc/[ddd]_{\TT(\rho_{\Gamma, A[\gamma]}^{-1})(\theta_{B, \gamma^+})\circ \theta_{A,
\gamma}^+}&
F\Gamma.FA[F\gamma]       
	\ar[r]^\p
	\ar@/^2pc/[ddd]^{\theta_{A, \gamma}}&
F\Gamma\\\\
&F\Gamma.F(A[\gamma]).\Pi(F(A[\gamma]),F(B[\gamma^+])[\rho_{\Gamma,A[\gamma]}^{-1}])[\p]
\ar@/_/[dl]_{\ev_{F(A[\gamma]),
F(B[\gamma^+])}}
        \ar[d]^\p
        \ar[r]^{\p^+}
        \pb{315}&
F\Gamma.\Pi(F(A[\gamma]),F(B[\gamma^+])[\rho_{\Gamma,A[\gamma]}^{-1}])       \ar[d]^{\p}\\
F\Gamma.F(A[\gamma]).F(B[\gamma^+])[\rho_{\Gamma, A[\gamma]}^{-1}]       \ar[r]^\p&
F\Gamma.F(A[\gamma])       \ar[r]^\p&
F\Gamma
}}
\]
where all arrows not explicitely displayed are identities.
\end{lem}
\begin{proof}
The only non-trivial equality to prove is that the morphism preserves evaluation, \emph{i.e.} that 
\[
\ev \circ \TT(\p)(\Pi(\theta_{A, \gamma}^{-1}, \TT(\rho_{\Gamma,
A[\gamma]}^{-1})(\theta_{B, \gamma^+})))\circ \theta_{A, \gamma}^+ = 
\TT(\rho_{\Gamma,
A[\gamma]}^{-1})(\theta_{B, \gamma^+})\circ \theta_{A,
\gamma}^+ \circ \ev
\]
which is a direct (if somewhat intricate) calculation on cwf combinators. Note that both $\ev$ are
evaluation morphisms for chosen dependent product diagrams, \emph{i.e.} $\ext{\p,
\app(\qI,\qI[\p])}$.
\end{proof}

Finally, the last thing we have to prove is that the upper morphism of the diagram of Lemma
\ref{lemma:piTh}, \emph{i.e.} $\theta_{\Pi(A,B), \gamma}$, induces as well a canonical morphism
between dependent product diagrams. 

\begin{lem}\label{lem:dia3}
The following is a morphism between two dependent product diagrams:
\[
\scalebox{0.8}{
\xymatrix@R=30pt{
&F\Gamma.FA[F\gamma].F(\Pi(A,B))[(F\gamma)\circ\p]      \ar@/_/[dl]_{\TT((F\gamma)^+)(\ev_{F(\Pi(A,
B))})}
        \ar@/^2pc/[ddd]^{\TT(\p)(\theta_{\Pi(A,B),\gamma})\circ \theta_{A,\gamma}^+}
        \ar[d]^\p
        \ar[r]^{\p^+}
        \pb{315}&
F\Gamma.F\Pi(A,B)[F\gamma]       
        \ar[d]^{\p}
        \ar@/^2pc/[ddd]^{\theta_{\Pi(A,B),\gamma}}\\
F\Gamma.FA[F\gamma].FB[\rho_{\Delta, A}^{-1}][(F\gamma)^+]       
	\ar[r]^\p
	\ar@/_2pc/[ddd]^{\TT(\rho_{\Gamma,A[\gamma]}^{-1})(\theta_{B,\gamma^+})\circ
\theta_{A,\gamma}^+}&
F\Gamma.FA[F\gamma]       
	\ar[r]^\p
	\ar@/_2pc/[ddd]^{\theta_{A,\gamma}}&
F\Gamma\\\\
&F\Gamma.F(A[\gamma]).F(\Pi(A,B)[\gamma])[\p]      \ar@/_/[dl]_{\ev_{F(\Pi(A[\gamma],B[\gamma^+]))}}
        \ar[d]^p
        \ar[r]^{\p^+}
        \pb{315}&
F\Gamma.F(\Pi(A,B)[\gamma])
        \ar[d]^\p\\
F\Gamma.F(A[\gamma]).F(B[\gamma^+])[\rho_{\Gamma, A[\gamma]}^{-1}]       \ar[r]^\p&
F\Gamma.F(A[\gamma])       \ar[r]^\p&
F\Gamma
}}
\]
where all arrows not explicitely displayed are identities.
\end{lem}
\begin{proof}
Recall that $\theta_{\Pi(A, B), \gamma}$ can be characterised as the unique morphism between two
candidate substitution pullbacks: one computed in $\C$ and transported via $F$, the other computed
in $\C'$. The proof that $\theta_{\Pi(A, B), \gamma}$ respects evaluation consists in redoing the
same reasoning, but with the whole dependent product diagram rather than just the type.

The following diagram represents the dependent product diagrams for $\Pi(A, B)$ and $\Pi(A[\gamma],
B[\gamma^+])$, along with the morphisms relating them together.

\[
\scalebox{.8}{
\xymatrix{
&&\Delta.A.\Pi(A,B)[\p]
	\ar@/_2pc/[ddll]
	\ar[dd]
	\ar[rr]&&
\Delta.\Pi(A,B)
	\ar[dd]\\
&&&\Gamma.A[\gamma].\Pi(A,B)[\gamma\circ\p]
	\ar[ul]
	\ar[dd]
	\ar[rr]
	\ar@/_2pc/[lldd]&&
\Gamma.\Pi(A,B)[\gamma]
	\ar[ul]
	\ar[dd]\\
\Delta.A.B
	\ar[rr]&&
\Delta.A
	\ar[rr]&&
\Delta\\
&\Gamma.A[\gamma].B[\gamma^+]
	\ar[rr]
	\ar[ul]&&
\Gamma.A[\gamma]
	\ar[rr]
	\ar[ul]&&
\Gamma	\ar[ul]
}}
\]
The front and back faces are both dependent product diagrams. We now map this diagram to $\C'$ via
$F$, and silently apply the canonical isomorphisms of the pseudo cwf-morphism structure, to obtain
(the bottom part of) the following diagram. We do not annotate the arrows to avoid cluttering the
diagram too much, but they can be recovered by carefully following the construction of the diagram.

\[
\scalebox{.65}{
\xymatrix@C=0pt{
&&&\hspace{-25pt}F\Gamma.FA[F\gamma].F(\Pi(A,B))[(F\gamma)\circ\p]
	\ar@/_4pc/[dddddll]
	\ar[ddl]
	\ar[rr]
	\ar@{.>}[ddd]|{\TT(\p)(\theta_{\Pi(A,B),\gamma})})
	\ar@/^2pc/[ddddd]&&
F\Gamma. F(\Pi(A,B))[F\gamma]
	\ar[ddl]
	\ar@/^2pc/[ddddd]
	\ar@{.>}[ddd]|{\theta_{\Pi(A,B),\gamma}}\\\\
&&F\Delta.FA.F(\Pi(A,B))[\p]
        \ar@/_2pc/[ddll]
        \ar[dd]
        \ar[rr]&&
F\Delta.F(\Pi(A,B))
        \ar[dd]\\
&&&F\Gamma.FA[F\gamma].F(\Pi(A,B)[\gamma])[\p]
        \ar[ul]
        \ar[dd]
        \ar[rr]
        \ar@/_2pc/[lldd]&&
F\Gamma.F(\Pi(A,B)[\gamma])
        \ar[ul]
        \ar[dd]\\
F\Delta.FA.FB[\rho_{\Delta,A}^{-1}]
        \ar[rr]&&
F\Delta.FA
        \ar[rr]&&
F\Delta\\
&\hspace{-25pt}F\Gamma.F(A[\gamma]).F(B[\gamma^+])[\rho_{\Gamma,A[\gamma]}^{-1}]
        \ar[rr]
        \ar[ul]&&
F\Gamma.F(A[\gamma])
        \ar[rr]
        \ar[ul]&&
F\Gamma  \ar[ul]
}}
\]

The top part of the diagram is obtained (up to an obvious isomorphism) by pulling back the dependent
product diagram in the back along $F\gamma$.
By the universal property of dependent products, the two morphisms from the top dependent product
diagram to the one in the back factor uniquely through the two dotted arrows. But for the right hand
side one, that exactly means that the condition of Lemma \ref{lem:theta_lemma} is satisfied and that
the right hand side dotted map is $\theta_{\Pi(A,B),\gamma}$. Similarly, the left hand side dotted
map is necessarily $\TT(\p)(\theta_{\Pi(A,B),\gamma})$. Therefore, it preserves the evaluation
maps, since it was constructed by the universal property of dependent product diagrams.

Annotating the morphisms following their construction, it becomes apparent that the commutation we
have proved is exactly the statement of the lemma.
\end{proof}

To wrap things up, we note that by the lemmas above all four morphisms of Lemma \ref{lemma:piTh}
correspond to canonical morphisms between dependent product diagrams: $i_{A[\gamma], B[\gamma^+]}$
by Lemma \ref{lem:dia0}, $\TT(F\gamma)(i_{A, B})$ by Lemma \ref{lem:dia1}, $\Pi(\theta_{A,
\gamma}^{-1}, \TT(\rho_{\Gamma, A[\gamma]}^{-1})(\theta_{B, \gamma^+}))$ by Lemma \ref{lem:dia2},
and $\theta_{\Pi(A,B),\gamma}$ by Lemma \ref{lem:dia3}. The corresponding morphisms between dependent product diagrams are composable,
and by uniqueness of the universal property it follows that the two paths of the diagram of Lemma
\ref{lemma:piTh} coincide.

\end{document}